\documentstyle[preprint,aps,prc,epsf,rotate]{revtex}
\begin{document}
\input{psfig}
\newcommand{\Del}{$\Delta$}
\newcommand{\g}{{\rm g}}
\def\beq{\begin{equation}}
\def\eeq{\end{equation}}
\def\bea{\begin{eqnarray}}
\def\eea{\end{eqnarray}}
\def\eqref#1{Eq.~(\ref{eq:#1})}
\def\eqlab#1{\label{eq:#1}}
\def\figref#1{Fig.~(\ref{fig:#1})}
\def\figlab#1{\label{fig:#1}}
\def\tabref#1{Table~\ref{tab:#1}}
\def\tablab#1{\label{tab:#1}}
\def\VYP#1#2#3{{\bf #1} (#2) #3}  
\def\NP#1#2#3{Nucl.~Phys. {\bf #1}, #3 (#2)}
\def\PLB#1#2#3{Phys.~Lett.~B {\bf #1}, #3 (#2)}
\def\PRC#1#2#3{Phys.~Rev.~C {\bf #1}, #3 (#2)}
\def\PR#1#2#3{Phys.~Rev. {\bf #1}, #3 (#2)}
\def\PRL#1#2#3{Phys.~Rev.~Lett. {\bf #1}, #3 (#2)}
\def\AP#1#2#3{Ann.~Phys.~(N.Y.) {\bf #1}, #3 (#2)}
\def\ZPA#1#2#3{Z.~Phys.~A {\bf #1}, #3 (#2)}
\newcommand{\thalf}{\mbox{\small{$\frac{3}{2}$}} }
\def\half{\mbox{\small{$\frac{1}{2}$}}}
\def\quarter{\mbox{\small{$\frac{1}{4}$}}}
\newcommand{\vslash}[1]{#1 \hspace{-0.5 em} /}
\draft
\tighten
\title{ Photoproduction of electron-positron pairs on the proton
in the resonance region }
\author{A.\ Yu.\ Korchin$^{a,b}$ and O.\ Scholten$^a$}
\address{$^a$ Kernfysisch Versneller Instituut, University of Groningen, 9747 AA Groningen, The~Netherlands}
\address{$^b$  National Science Center `Kharkov Institute of Physics and Technology',
310108 Kharkov, Ukraine}
\maketitle
\begin{abstract}
Production of lepton pairs in the $\,\gamma p \rightarrow e^-e^+ p\,$ reaction
is studied at photon energies up to 1 GeV.
We show that even if lepton charge is not measured there are extensive kinematical regimes where
the nuclear process dominates over the Bethe-Heitler contribution.
The decomposition is performed of the unpolarized cross section for the virtual Compton scattering
in terms of response functions.
These are expressed in terms of a polarization density matrix of the virtual photon
and are shown to be sensitive to the properties of
baryon resonances in the first and the second resonance regions.
In the analysis a unitary K-matrix
model is used, based on an effective Lagrangian including
nucleon, mesons, and baryon resonances with masses up to 1.7 GeV.
Results of the model are compared with data for real-photon Compton scattering
and predictions are given for observables in photoproduction of lepton pairs.
\end{abstract}
\bigskip
\noindent
{\bf 1996 PACS} numbers: 13.30.-a, 13.40.-f, 13.40.Gp, 13.40.Hq, 13.60.Fz\\
{\bf Key Words}  Virtual Compton scattering;  Real Compton scattering;  Baryon resonances;
K-matrix approach \\

\section{Introduction}  \label{sect:intro}

Compton scattering (CS) on the proton with real and virtual photons is
a fundamental process which provides information
on the internal structure of the proton and its excited states.
Virtual CS where the initial photon is space-like, i.e. the reaction $e^- p\to e^- p\gamma\,$,
has recently attracted
considerable attention. Below the pion production threshold it allows,
via an interference of the Bethe-Heitler (BH) process with the proton virtual CS,
to measure the generalized polarizabilities of the proton.
For a review we refer to~\cite{Procee,Gui98}.

Virtual CS in the time-like region,
the reaction $\,\gamma p\to e^- e^+ p\,$, is of interest for complementarity reasons.
The process is sensitive to the nucleon electromagnetic (e.m.) form factors
in the region $0 < k^2 < 4m^2$ (where $m$ is the nucleon mass, $k$ is the photon momentum), which
are not accessible in elastic electron-proton scattering or
$e^+ e^-$ annihilation to $p\bar{p}\,$. This possibility was explored in Ref.~\cite{Sch95}.
In the present paper the main interest is to investigate
dependence of the polarization density matrix of the virtual
photon on various contributions to the nuclear matrix element.

In certain parts of phase space the cross section is dominated by the BH process.
Only  when lepton charges are measured this can be turned to an advantage by measuring
an $e^+ e^-$ asymmetry which is directly proportional to the virtual CS -- BH interference.
This was suggested long ago~\cite{Alv73} and more recently elaborated in~\cite{Lvo96,Kor97,Die97}.

In this paper we study the situation where the electron and positron in
the $\gamma p\to e^- e^+ p$ reaction are not distinguished which is the case for
experiments which are being done at MAMI (Mainz)~\cite{Bac99}.
Under such conditions, because of the different charge-conjugation parity of virtual
CS and BH amplitudes~\cite{Bjo58},
the above interference vanishes and one is left with the incoherent sum
of virtual CS and BH contributions.
We show that there are large kinematical regimes where the more interesting virtual CS on the
proton is dominant, mainly at backward angles for the virtual photon.
In these conditions the (unpolarized) $\gamma N \to e^- e^+ N$
cross section can be decomposed in terms of response functions (RF)s,
like is usually done for exclusive electron scattering on nuclei (see, e.g.~\cite{Don86}).
In addition to the transverse RF, which can also be measured in unpolarized CS with real photons,
there are three more RFs for dilepton production.
These are directly related to polarization properties of the virtual photon. To make this link
more transparent we apply to the $\gamma N \to \gamma^* N$ reaction the density matrix formalism
developed~\cite{Sch70} for the photoproduction of vector mesons on the proton.

Differential cross sections and RFs for the $\gamma p \to e^- e^+ p$ reaction are
calculated in the unitary K-matrix
model previously developed~\cite{Sch96,Kor98} for pion-nucleon scattering, pion photoproduction and
real-photon CS. Before applying the model to virtual CS we compare its predictions to data
for real CS in the energy region up to 1 GeV.

In the calculation of virtual CS we concentrate on the 1st and the 2nd resonance
regions which are clearly seen in the cross section for real CS. The cross section
and elements of the polarization density matrix
$\rho_{\lambda \lambda'}$ are calculated for two energies
corresponding to excitation of the  $P_{33}(1232)$ and the $D_{13}(1520)$ resonances.
We study effects of different reaction mechanisms on these observables.
In particular, for the \Del\ region it is of interest to investigate the effect of
t-channel exchange of the $\sigma$ meson.

At photon energies of about 700 MeV photon invariant masses
up to $\approx$ 540 MeV become accessible.
In this energy regime several contributions are important, in particular,
the $D_{13}(1520)$ and $S_{11}(1535)$ resonances, and $\sigma$ exchange.
Due to their different influence on the density matrix elements
$\rho_{\lambda \lambda'}$ these contributions can be distinguished
in an experiment provided RFs are extracted.
Cross section and other observables are also sensitive to the $G_{3}(k^2)$ coupling in the
$\gamma N D_{13}(1520)$ vertex
which is specific for virtual photon and does not contribute to real CS.

The paper is organized as follows. In sect.~\ref{sec:formalism} we specify
variables and give the expression for the
exclusive cross section. The decomposition of the virtual CS cross section in terms of RFs is
presented. Relations between RFs and polarization density matrix elements are established and
the limit of ``almost'' real photons is considered.
We briefly discuss ingredients of the dynamical model used and
compare calculated cross sections and polarization observables with data for real CS.
In sect.~\ref{sect:results} kinematical regimes are investigated where the BH contribution can be
neglected compared to virtual CS.
Results for response $W_S =W_T +W_L$ and elements of the photon density matrix are presented.
Conclusions are given in sect.~\ref{sect:conclusions}.
In Appendix A  expressions for RFs for virtual CS and polarization observables for real CS
in terms of helicity amplitudes are collected.

\section{Formalism} \label{sec:formalism}

\subsection{Cross section for $\gamma p\rightarrow e^- e^+ p$ reaction}   \label{subsec:crosssection}

If $q=(q_0 =|\vec{q}| ,\vec{q})$ is the momentum of the incoming real photon
and $k= (k_0 ,\vec{k})=k_- + k_+$
the momentum of the outgoing virtual photon (see~\figref{figu:1}),
the fully exclusive cross section in the c.m. frame can be written as~\footnote{We follow notation
of Ref.~\cite{Itz80} }
\bea
\frac{d\sigma (e^- e^+ )}{dM_\gamma d \Omega_\gamma d \Omega_l}=
{\cal K}_{e^- e^+} |A_{BH}+A_{VCS}|^2\,,
\eqlab{crosssec}
\eea
where $M_\gamma = \sqrt{k^2}$ is the photon invariant mass
and ${\cal K}_{e^- e^+}$ is the kinematical factor
\bea
{\cal K}_{e^- e^+} = \frac{m^2 m_{e}^2}{2(2\pi)^5 s}\frac{|\vec{k}|}{|\vec{q}|}\frac{4 |\vec{l}|^3 }
{M_\gamma \beta^2 k_0 }\,.
\eqlab{kinemat}
\eea
Here $m (m_e )$ is the proton (electron) mass, $\beta=(1-4m_e ^2 / M_\gamma ^2)^{1/2}$ is the $e^-$ ($e^+$)
velocity in the virtual-photon rest frame, and $s$ is the invariant energy squared.
The average over initial and sum over final polarizations of all particles is implied in \eqref{crosssec}
as well as in the following.

To describe the $e^-e^+$
pair the relative 4-momentum $l =\frac{1}{2}(k_- -k_+ )$ has been
introduced~\cite{Nec94}. The orientation of $\,\vec{l}\,$
is determined by the polar angle $\theta_l$ and the azimuthal (or out-of-plane) angle $\phi_l$ which are
defined in the frame with $OZ$ axis along $\vec{k}$ and $OY$ axis along $\vec{k} \times \vec{q}$.
The magnitude of $\vec{l}$ is given by
\beq
|\vec{l}| = \frac{M_\gamma \beta k_0 }{ 2 (M_\gamma^2 +\vec{k}^2 \sin^2 {\theta_l} )^{1/2} }
\eqlab{rel}
\eeq
with $k_0 = (s-m^2 +M_\gamma^2 )/2\sqrt{s}$ and $\vec{k}^2 = k_0^2-M_\gamma^2$.
The solid angle differential $d\Omega_l$ stands for $d \cos{\theta_l} d \phi_l$, and
$d\Omega_\gamma=d \cos{\theta_\gamma} d\phi_\gamma$,  where $\theta_\gamma$ is the angle between vectors
$\vec{k}$ and $\vec{q}$. The azimuthal angle $\phi_\gamma$ is superfluous and can be chosen zero.
For more details about kinematics we refer to previous papers on $e^- e^+$ production
in capture reactions~\cite{Nec94,Kor98a} and $NN$ bremsstrahlung~\cite{Kor95}.

The Bethe-Heitler ($A_{BH}$) and the virtual CS ($A_{VCS}$) amplitudes are functions of five variables:
$s, M_\gamma ,\theta_\gamma , \theta_l$ and $ \phi_l$. Before proceeding further one can exploit
the different charge-conjugation properties of these amplitudes~\cite{Bjo58}.
Interchanging variables describing $e^-$ and $e^+$, which means
$\vec{l}\rightarrow -\vec{l}\,$ and interchange $e^-$ and $e^+$ helicities,
leads to the relations
\bea
&&|A_{BH}(\theta_l,\phi_l )|^2=|A_{BH}(\pi -\theta_l,\pi+ \phi_l )|^2\,, \nonumber\\
&&|A_{VCS}(\theta_l,\phi_l )|^2=|A_{VCS}(\pi -\theta_l,\pi+ \phi_l )|^2\,, \nonumber\\
&&A_{BH}^*(\theta_l,\phi_l )A_{VCS}(\theta_l,\phi_l )
= - A_{BH}^*(\pi -\theta_l,\pi+ \phi_l )A_{VCS}(\pi-\theta_l,\pi+\phi_l )\,,
\eea
where $\theta_l$ varies from $0^\circ$ to $180^\circ$ and only the dilepton angles are indicated.
Therefore, if the electron and positron cannot be distinguished in an experiment,
 the cross section is the incoherent sum
$d \bar{\sigma} (e^- e^+ )= d\sigma(e^- e^+ ) +d\sigma(e^+ e^- )\,$, which results in
\bea
\frac{d\bar{\sigma}(e^- e^+ ) }{dM_\gamma d \Omega_\gamma d \Omega_l}=
2{\cal K}_{e^- e^+} (\, |A_{BH}|^2 +|A_{VCS}|^2\, ) \,,
\eqlab{defcs}
\eea
where the interference term drops out.

Another interesting observable, the $e^- e^+$ asymmetry,
which is proportional to the virtual CS -- BH interference (see~\cite{Alv73,Lvo96,Kor97,Die97}),
will not be addressed here. Measuring this would require detection of the charges of the leptons.

The expression for $A_{BH}$ can be read off the diagrams g,h in~\figref{figu:1}. After some algebra it
takes the form
\bea
A_{BH}= \frac{e^3}{Q^2}
\bar{u}(k_- ) [ \,&& \gamma_\mu ( \frac{k_+ ^\nu}{k_+ \cdot q} -  \frac{k_- ^\nu}{k_- \cdot q})
+\frac{\gamma^\nu \vslash{q}\gamma_\mu }{2k_-\cdot q}-
\frac{\gamma_\mu \vslash{q} \gamma^\nu }{2k_+ \cdot q}\,]  v(k_+ ) \nonumber\\
&&\times \bar{u}(p' ,\Lambda')[F_1 (Q^2 )\gamma^\mu +i\frac{\sigma^{\mu\rho}Q_\rho}{2m}F_2 (Q^2)]u(p ,\Lambda)
\epsilon_\nu (\lambda)\,,
\eqlab{BH}
\eea
where $u(p ,\Lambda)$ ($\bar{u}(p' ,\Lambda')$) is the spinor of the initial (final) nucleon with
momentum $p$ ($p'$) and helicity $\Lambda$ ($\Lambda'$).
In \eqref{BH} $e$ is the proton charge, $\bar{u}(k_- )$ and $ v(k_+ )$ are the lepton spinors (omitting
spin indices), and $\epsilon(\lambda)$ is the photon polarization vector for helicity $\lambda$.
The nucleon e.m.\ form factors $F_{1,2}(Q^2 )$ depend on the momentum $Q=p'-p$.

The virtual CS matrix element is
\bea
A_{VCS} = \frac{e^2}{M_\gamma^2} \bar{u}(k_- )\gamma_\mu v(k_+ ) J^\mu\,,
\eea
where the e.m.\ current $J^\mu$ is expressed through the CS tensor $M^{\mu\nu}$
\bea
J^\mu = e\bar{u}(p' ,\Lambda')M^{\mu\nu}u(p ,\Lambda)\epsilon_\nu(\lambda)\,.
\eea
Some of the ingredients in the calculation of  $M^{\mu\nu}$ are discussed in
sect.~\ref{subsec:parameters}.

Note that the real CS matrix element ($A_{RCS}$) can be written in this notation as
\bea
A_{RCS} = e^2 \epsilon^{'*}_\mu (\lambda') \bar{u}(p' ,\Lambda')M^{\mu\nu}u(p ,\Lambda)\epsilon_\nu(\lambda)
\eea
with $\epsilon'(\lambda')$ being the transverse polarization vector of the final photon
which is real in this case.

\subsection{Response functions and polarization density matrix }
\label{subsec:response}

In kinematics where virtual CS is the dominant process it is of interest to
decompose the cross  section
in terms of response functions (RF)s, similar to what is common practice in the space-like region.
We will not go in much detail because the derivation is similar to $e^- e^+ $ production
in other reactions~\cite{Nec94,Kor98a,Kor95}.

After summing over lepton polarizations and making use of gauge invariance $k\cdot J =0$
one obtains in the c.m. frame
\bea
&&\frac{d\bar{\sigma}(e^- e^+ ) }{dM_\gamma d \Omega_\gamma d \Omega_l}=
\frac{\alpha^2 m^2 |\vec{l}|^3}{2\pi^3 s M_\gamma^3\beta^2 k_0}\frac{|\vec{k}|}{|\vec{q}|}
\,2 {\cal S}\,,
\eqlab{ltcross} \\
&& {\cal S} =
 W_T x_T + W_L x_L +W_{TT} x_{TT}\cos{2\phi_l}+W_{LT} x_{LT} \cos{\phi_l} \,,
\eqlab{lt}
\eea
where  $\alpha$ is the fine-structure constant,
$\,W_i \equiv W_i (s,\theta_\gamma ,M_\gamma )\,$ are RFs for $\,i=T,L,TT\,$ and $\,LT\,$, and
the factor $2$ in front of ${\cal S}$ comes from the definition \eqref{defcs}.
Dependence on the polar angle $\theta_l$ is contained in the factors $x_i$
\bea
&& x_T =1-2\frac{\vec{l}^2}{M_\gamma^2}\sin^2{\theta_l}\,,\;\;\;\;\;\;\;\;\;\;\;
x_L =1-4\frac{\vec{l}^2}{k_0^2}\cos^2{\theta_l} \,,\nonumber\\
&& x_{TT} =2\frac{\vec{l}^2}{M_\gamma^2}\sin^2{\theta_l}\,,\;\;\;\;\;\;\;\;\;\;\;
x_{LT} =\sqrt{2}\frac{\vec{l}^2}{M_\gamma k_0}\sin{2\theta_l}\,,
\eqlab{xfactors}
\eea
where $|\vec{l}|$, which also depends on $\theta_l$, is defined in \eqref{rel}.

RFs are defined in terms of the space components $J^i$ of the e.m.\ current as~\cite{Kor98a}
\bea
&&W_T =\frac{1}{4}\sum_{polar.}|J_x |^2 +|J_y |^2\,,\;\;\;\;\;\;\;\;\;\;\;
W_L = \frac{1}{4} \frac{k^2}{k_0^2}\sum_{polar.}|J_z |^2 \,,\nonumber\\
&&W_{TT} =\frac{1}{4} \sum_{polar.}|J_y |^2 -|J_x |^2\,,\;\;\;\;\;\;\;\;
W_{LT} = -\frac{1}{4}\frac{ \sqrt{k^2} }{k_0} \sum_{polar.} 2 \sqrt{2}\Re\, (J_z^* J_x )\,,
\eqlab{response}
\eea
where the sum runs over the proton and initial photon polarizations.

It is interesting to note that certain combinations of the e.m.\ current vanish
due to reflection symmetry ($OY \to -OY$) with respect to the scattering plane $OXZ$:
\bea
\sum_{polar.} \Re\, (J_x^* J_y )= \sum_{polar.} \Im\, (J_x^* J_y )=
\sum_{polar.} \Re \, (J_z^* J_y ) = \sum_{polar.} \Im\, (J_z^* J_y ) =0\,.
\eea
These constraints are specific for a two-body final state such as $N\,+ \gamma^*$
and may be useful for checking consistency of a model.

Experimental separation of RFs can be performed using the different dependencies of the kinematical factors
in \eqref{lt} on the dilepton angles. In Refs.~\cite{Mes} an example of such a separation for the case of
$NN$ virtual bremsstrahlung is discussed.

To obtain some further insight we relate the above RFs to elements of the polarization density
matrix $\hat{\rho}$ of the virtual photon.
The formalism of the polarization density matrix has been used before in
production of vector mesons $\rho, \omega$ (see Refs.~\cite{Dab67,Sch70,Klo98,Klo99}).
One introduces the $3 \times 3$ density matrix for
a spin-1 particle~\cite{Sch70}
\bea
\hat{\rho} &=& \left( \begin{array}{ccc}
\rho_{11}& \rho_{10} &\rho_{1-1} \\
\rho_{10}^*              &\rho_{00}  &-\rho_{10}^* \\
\rho_{1-1}                &-\rho_{10}  & \rho_{11}
        \end{array}\right)\,,
\eqlab{density1}
\eea
in the helicity basis $\lambda, \lambda' = 0,\pm 1$. It
satisfies the conditions of hermiticity $\rho_{\lambda\lambda'} =\rho^*_{\lambda'\lambda}$,
parity conservation  $\rho_{\lambda\lambda'} =(-1)^{\lambda-\lambda'} \rho_{-\lambda -\lambda'}$, and has
a unit trace ($\rho_{11} = \frac{1}{2}(1-\rho_{00})$).
$\rho_{00}$ and $\rho_{1-1}$ are thus real quantities while the element $\rho_{10}$ is, in general, complex.

The form \eqref{density1} is valid only for a reaction with a two-body final
state (like $N+ \gamma^*$), in which the initial hadron and photon are not polarized and
the polarization of the final hadron is not measured. If the initial photon in $\gamma N \to \gamma^* N$
is (fully or partially) polarized  the general density matrix is more complicated~\cite{Sch70}.

From the density matrix and the $\,\,\gamma e^- e^+\,$ vertex, which is well-known
from QED,
one can obtain the angular distribution $F (\bar{\theta}_l ,\bar{\phi}_l )$
of $\gamma^*$ decay into $e^- e^+$.
In the virtual-photon rest frame  it takes the form
(see Ref.~\cite{Pilk79}, ch. 4-13~\footnote{There is a misprint in Eq.(13.14) of this reference:
in front of $a_1$ there should be an additional factor of 1/2 } )
\bea
F (\bar{\theta}_l ,\bar{\phi}_l ) =\frac{1}{4\pi}\, \{\, 1-\frac{\beta^2}{3-\beta^2}
\,[\, && \frac{1}{2} (3\rho_{00}-1)(3\cos^2 \bar{\theta}_l -1)\nonumber\\
&& -3\rho_{1-1}\sin^2 \bar{\theta}_l \cos 2\bar{\phi}_l -3\sqrt{2} \Re \rho_{10} \sin 2\bar{\theta}_l
\cos \bar{\phi}_l \,]\, \}\, ,
\eqlab{angular1}
\eea
where $\beta$ is defined after \eqref{kinemat} and  $( \bar{\theta}_l,\,\bar{\phi}_l )$
denote the dilepton angles.

Expression \eqref{lt} in terms of RFs, boosted to the $\gamma^*$
rest frame, can be cast in the form (details on the
angle transformation can be found in Refs.~\cite{Nec94,Kor95})
\bea
{\cal S}&=&
 W_T (1-\frac{1}{2}\beta^2 \sin^2 \bar{\theta}_l ) + W_L (1- \beta^2 \cos^2 \bar{\theta}_l )
 +W_{TT} \frac{1}{2} \beta^2 \sin^2 \bar{\theta}_l \cos 2\bar{\phi}_l \nonumber\\
 &+& W_{LT} \frac{1}{2\sqrt{2}}\beta^2 \sin 2\bar{\theta}_l \cos \bar{\phi}_l   \nonumber\\
 &=& (1-\frac{\beta^2}{3})W_S \,  \{
1-\frac{\beta^2}{3-\beta^2}\, \big[ \,
 \frac{2W_L-W_T}{2W_S} (3\cos^2 \bar{\theta}_l -1 )
 - \frac{3W_{TT}}{2W_S} \sin^2 \bar{\theta}_l \cos 2\bar{\phi}_l \nonumber\\
 &-& \frac{3W_{LT}}{2\sqrt{2} W_S} \sin 2\bar{\theta}_l \cos \bar{\phi}_l\, \big]\, \}\,,
\eqlab{angular3}
\eea
where $W_S \equiv W_T + W_L\,$.
It is seen that the angle dependent part of ${\cal S}$
has the same structure as $F (\bar{\theta}_l ,\bar{\phi}_l )$ in \eqref{angular1}.
Comparing terms with the same angle factors one obtains
\bea
\rho_{00}=\frac{W_L}{W_S}\,,\;\;\;\;\;\;\;\;\rho_{1-1}=\frac{W_{TT}}{2 W_S}\,,\;\;\;\;\;\;\;\;
 \Re \rho_{10} =\frac{W_{LT}}{4 W_S}\,.
\eqlab{rho-w}
\eea

Note that the imaginary part of $\rho_{10}$ does not enter in the $\gamma^* \to
e^- e^+$  angular
distribution (cf.~\cite{Sch70} for $\gamma N \to \rho N \to \pi \pi N$).
For completeness we give an expression for $\Im \rho_{10}$ through an additional (independent) function
$\tilde{W}_{LT}$:
\bea
\Im \rho_{10} =\frac{\tilde{W}_{LT}}{4 W_S}\,,\;\;\;\;\;\;\;\;\;\;\;\;\;\;\;
\tilde{W}_{LT} = -\frac{1}{4}\frac{ \sqrt{k^2} }{k_0} \sum_{polar.} 2 \sqrt{2}\Im\, (J_z^* J_x )\,.
\eqlab{addit-LT}
\eea
${W}_{LT}$ and  $\tilde{W}_{LT}$ can be regarded as the real and the imaginary parts of one complex function.
Thus only three (out of four independent) components of the photon-polarization density matrix in
addition to the cross section can be determined from measuring
$W_T, W_L, W_{TT}$, and $W_{LT}$ in the $\gamma N \to e^- e^+ N$ reaction with {\it unpolarized leptons}.

Since $|W_{TT}| \le W_T \le W_S $ (see definitions in Eqs.~(\ref{eq:response})), the following inequalities
are in order
\bea
0 \le \rho_{00} \le 1\,,\;\;\;\;\;\;\;\;\;
|\rho_{1-1}| \le \frac{1}{2} (1- \rho_{00}) \le \frac{1}{2}\,.
\eea
The upper bound for $\rho_{10}$ was obtained in  Ref.~\cite{Dab67} using the Schwartz inequality
\bea
|\rho_{10}| \le  \frac{1}{2} \sqrt{\rho_{00} (1 - \rho_{00} -2 \rho_{1-1}) }
\le \frac{1}{\sqrt{2}} \sqrt{ \rho_{00} (1 - \rho_{00}) } \le \frac{1}{2\sqrt{2}}\,.
\eea
Additional constraints have recently been derived in~\cite{Klo99}.

Sometimes (see~\cite{Robs74,Klo98}) the density matrix for the spin-1 particle is expressed
in terms of tensor and vector polarizations
\bea
\hat{\rho} &=&  \frac{1}{3}\left( \begin{array}{ccc}
1+ \frac{1}{\sqrt{2}}t_{20} & \sqrt{\frac{3}{2}}t_{1 1}-  \sqrt{\frac{3}{2}}t_{21}  &\sqrt{3}t_{22} \\
-\sqrt{\frac{3}{2}}t_{11}-\sqrt{\frac{3}{2}}t_{21}&1- \sqrt{2}t_{20}&\sqrt{\frac{3}{2}}t_{11}+
\sqrt{\frac{3}{2}}t_{21} \\
\sqrt{3}t_{22}  & -\sqrt{\frac{3}{2}}t_{11}+ \sqrt{\frac{3}{2}}t_{21}  &1+ \frac{1}{\sqrt{2}}t_{20}
        \end{array}\right)\,,
\eqlab{density2}
\eea
where the tensor polarizations $t_{20},t_{21}$ and $t_{22}$ are real, and $t_{11}$ is purely
imaginary and related to the vector polarization $p_y$ via $\,t_{11}=-i\frac{\sqrt{3}}{2} p_y$.
Relations between these polarizations and elements $\rho_{\lambda \lambda'}$ follow from
\eqref{density1} and \eqref{density2},
\bea
\rho_{00}=\frac{1}{3}(1-\sqrt{2}t_{20})\,,\;\;\;\;\;\;\;
\rho_{1-1}=\frac{1}{\sqrt{3}}t_{22}\,,\;\;\;\;\;\;\;
\Re \rho_{10}=-\frac{1}{\sqrt{6}}t_{21}\,,\;\;\;\;\;\;\;
\Im \rho_{10}=-\frac{i}{\sqrt{6}}t_{11}\,.
\eqlab{rho-t}
\eea

Using Eqs.~(\ref{eq:rho-w}) and Eqs.~(\ref{eq:rho-t}) one can express the polarizations in terms of RFs
\bea
&&t_{20}= \frac{1}{\sqrt{2}}\frac{W_T -2 W_L}{W_S}\,, \;\;\;\;\;\;\;\;\;\;\;\;\;\;\;
  t_{22}= \frac{\sqrt{3}}{2} \frac{W_{TT}}{W_S}\,,       \nonumber\\
&&t_{21}= - \frac{\sqrt{3}}{2\sqrt{2}} \frac{W_{LT}}{W_S}\,,\;\;\;\;\;\;\;\;\;\;\;\;\;\;\;\;\;
 t_{11}= i \frac{\sqrt{3}}{2\sqrt{2}} \frac{\tilde{W}_{LT}}{W_S}\,.
\eqlab{t-w}
\eea

As a last point it is interesting to consider the limiting case of small invariant masses $M_\gamma$
close to the threshold value $M_\gamma^{th}=2m_e$. In this case the longitudinal
component of the e.m.\ current
becomes negligibly small, and elements $\rho_{00}$ and $\rho_{10}$ vanish.
As follows from Eqs.~(\ref{eq:t-w}), $\,t_{20} \to 1/ \sqrt{2},\; t_{21},\, t_{11} \to 0$.
In this limit \eqref{density1} reduces to
\bea
\hat{\rho} &=& \frac{1}{2} \left( \begin{array}{ccc}
1 & 0 & \frac{W_{TT}}{W_T} \\
0 & 0  & 0 \\
 \frac{W_{TT}} {W_T} & 0  & 1
        \end{array}\right)\,.
\eqlab{density3}
\eea
From a physical point of view this limit is close to real CS.
A real photon is, in general, described by the $2\times 2$ density matrix in the helicity
basis~\cite{Sch70}
\bea
\hat{\rho}_\gamma &=& \frac{1}{2} \left( \begin{array}{cc}
1 \pm I_C  &  - I_L \exp(-2 i\psi ) \\
-I_L \exp(2 i\psi )  & 1 \mp I_C
        \end{array}\right)\,,
\eqlab{real-dens}
\eea
where $I_L (I_C )$ is the degree of linear (circular) polarization, $\psi$ is the angle between
the direction of linear polarization and the $OX$ axis, and $+$ ($-$) stands for the right (left)
circular polarization.
Comparing \eqref{real-dens} with \eqref{density3} (dropping there the superfluous line and column
with elements equal to zero) one finds
\bea
I_C  &=& 0\,,\;\;\;\;\;\;\;\; I_L = \frac{|W_{TT}|}{W_T}\,<\,1, \nonumber\\
\psi\;\; &=&\;\;\Big\{ \;\;
\begin{array}{cc}
0\,,             & \text{ \hspace{0.2cm} if \hspace{0.4cm}  sign($W_{TT}) = -1$ } \,,\\
\frac{\pi}{2}\,,         & \text{ \hspace{0.2cm} if \hspace{0.4cm}  sign($W_{TT}) = +1$ } \,.
\end{array}
\eea
Thus the ``quasi-real" photon has no circular polarization, and may have a linear
polarization along the $OX$ axis (in the scattering plane), or along the $OY$ axis
(perpendicular to the scattering plane) depending on the sign of the interference $W_{TT}$.
Note that this is a particular feature of any reaction with unpolarized particles in coplanar kinematics
and can be understood based on reflection symmetry with respect to the scattering plane.

The photon asymmetry~\cite{Pfe74}, as can be measured in real CS with a
linearly-polarized photon beam, is related to
$W_T^r$ and $W_{TT}^r$ (the superscript ``r" denotes taking the real CS limit).
If there are no bound states in the energy regime of interest,
the RFs for $e^- e^+$ production behave smoothly as functions of
$M_\gamma$~\footnote{In principle, kinematical conditions exist
when the electron interacts strongly with the proton via
the Coulomb attraction and may be captured in
a hydrogen atom. However, this happens in a tiny region of the phase space and
is neglected in the present model. Likewise  the Coulomb interaction between the $e^-$ and the
$e^+$ and possible formation of positronium is not considered}
and the threshold values $W_{T,\,TT}$ at $M_\gamma = 2m_e $ can be used to calculate $W^r_{T,\,TT}$.
Making use of \eqref{real-dens} and time-reversal invariance one obtains for the real CS cross section
\bea
\frac{d\sigma (\gamma)}{d\Omega_\gamma}=
\frac{\alpha m^2}{4\pi s}\,(W^r_T - W^r_{TT} I_L \cos{2\psi})\,,
\eqlab{realcs}
\eea
where the angle $\psi$ specifies the direction of the incoming photon polarization
and all other particles are not polarized.
The cross section for unpolarized photons is expressed via $W^r_T$, and the
asymmetry is proportional to the interference $W^r_{TT}$:
\bea
\Sigma_\gamma = \frac{d\,\sigma_{\perp}(\gamma) -d\,\sigma_{\parallel}(\gamma)}
{ d\,\sigma_{\perp} (\gamma)
+ d\,\sigma_{\parallel}(\gamma) }\, = \,\frac{ W^r_{TT} }{W^r_T }\,.
\eqlab{asymmetry}
\eea

Calculations of the cross section and RFs for virtual and real CS are performed in the helicity formalism
(see Appendix A where details are presented).

 It is of interest to make a link between the representation in \eqref{lt} and the longitudinal--transverse
 (L--T) decomposition established for electron scattering on nuclei.
 At invariant masses $M_\gamma$ considerably larger than $2m_e$ one can rewrite \eqref{lt} as
 \bea
 {\cal S}=
 2x_T\,[\,(\frac{1}{2} W_T ) - \varepsilon_+ \,(-W_L)
 &-& \varepsilon_+  \cos{2\phi_l}\, (- \frac{1}{2} W_{TT}) \nonumber\\
  &+& \eta \sqrt{2\varepsilon_+ (1-\varepsilon_+ )} \cos{\phi_l}\, (\frac{1}{2\sqrt{2}} W_{LT}) \,]\,,
 \eqlab{newform}
 \eea
 where $\eta =\,\mbox{sign}\,(\sin2\theta_l )$, and
 $\varepsilon_+ =(1-x_T )/x_ T$ which satisfies the conditions $0\,<\varepsilon_+ \,<1$.
 Expression~\eqref{newform} is similar to the $L-T$ decomposition
 used in electron scattering (see e.g.,\cite{Don86,Dre92})
 To obtain exactly the same form  one can make in \eqref{newform} the formal
 substitutions $k^2 \to -|k^2|$ and $\varepsilon_+ \to  -\varepsilon_-$. This results in
 \bea
 {\cal S}=
 2x_T\,[\,R_T + \varepsilon_- \, R_L
 + \varepsilon_-  \cos{2\phi_l}\, R_{TT}
 + \eta \sqrt{2\varepsilon_- (1+\varepsilon_- )} \cos{\phi_l}\, R_{LT} \,]\,,
 \eqlab{newform1}
 \eea
  where RFs for the electron scattering, denoted by $R_i$, are related to $W_i$ via
 \bea
  R_T =\frac{1}{2} W_T\,,\;\;\;\;\;\;\;\;\;\;\;\; R_L = - W_L\,,\;\;\;\;\;\;\;\;\;\;\;\;
 R_{TT}= - \frac{1}{2} W_{TT}\,,\;\;\;\;\;\;\;\;\;\;\;\; R_{LT}=-\frac{i}{2\sqrt{2}} W_{LT}\,.
 \eqlab{electron}
 \eea
 Comparison of \eqref{newform1} with the $L-T$ decomposition in~\cite{Dre92} (Eqs.(22,23))
 suggests that $\varepsilon_-$ has a meaning of degree of transverse polarization
 of the space-like photon. The interference $R_{LT}$ is of course a real function in this case
 since the factor $i$
 merely accounts for $\sqrt{k^2} \to i \sqrt{\vert k^2 \vert}$ in the definition of Eqs.~(\ref{eq:response}).

\subsection{Description of the model and comparison with data for real Compton scattering}
\label{subsec:parameters}

As mentioned in sect.~\ref{sect:intro} we employ the K-matrix model
which is unitary in the channel space $\pi N \oplus \gamma N$.
The tree-level K-matrix includes
the four-star baryon resonances: P$_{11}$(N1440), D$_{13}$(N1520), S$_{11}$(N1535), S$_{11}$(N1650),
P$_{33}$($\Delta$1232), S$_{31}$($\Delta$1620) and D$_{33}$($\Delta$1700)
in the s- and u-type diagrams.
The t-channel contributions come from the exchange of $\sigma$ and
$\rho$ mesons (for $\pi N$ scattering), $\pi$, $\rho$, and  $\omega$
mesons (for pion photoproduction), and $\pi^0$, $\eta$ and  $\sigma$ mesons (for CS).
The amplitude for CS is shown in~\figref{figu:1}, diagrams a -- f.
For more details about the model we refer to Ref.~\cite{Kor98}.
In order to calculate the virtual CS the longitudinally polarized photons
have also been taken into account.

In this paper we concentrate on the $P_{33}(1232)$ and the $D_{13}(1520)$ resonance regions.
The $\gamma N R$ vertex  for the e.m.\ couplings of the spin-3/2\ resonances
is chosen as follows~\cite{Pec69,Ols75,Ben89,Pas95}
\bea
&&V_{N\gamma \rightarrow R^\alpha}= \frac{ie}{2m}\,
[\,G_1(k^2)\theta_{\alpha\beta}(z_1)\gamma_\delta
-\frac{G_2 (k^2 )}{2m}\theta_{\alpha\beta}(z_2 )p_\delta
-\frac{G_3 (k^2 )}{2m}\theta_{\alpha\beta}(z_3) k_\delta \,]\,  \Gamma\,
(k^\beta\epsilon^\delta -k^\delta\epsilon^\beta)\,,
\eqlab{vdelta}
\eea
and $\,V_{R^\alpha \rightarrow N\gamma}= - \gamma_0 V_{N\gamma \rightarrow R^\alpha}^\dagger \gamma_0\,$.
In \eqref{vdelta} $p\,(k)$ is the nucleon (photon) momentum and $\epsilon$ is the photon
polarization vector. For isospin-1/2\ resonances
$\, G_i (k^2)= [\, g_{i}^{(p)} (1+\tau_3 )/2 + g_{i}^{(n)} (1-\tau_3 )/2 \, ] F_{VMD}(k^2 )\,$, while
for isospin-3/2\ case $G_i (k^2 )= g_{i} T_3 F_{VMD}(k^2 )\,$, where
$T_3$ is $1/2 \leftrightarrow 3/2$ isospin-transition operator and
$\,F_{VMD}(k^2)\,$ is the form factor in the extended VMD model~\cite{VMD}.
Note that the $G_3(k^2)$ term contributes only for virtual photons.
Further, $\,\Gamma=(\gamma_5\,,1)$ for the resonances (3/2$^+$, 3/2$^-$), and the
tensor $\,\theta_{\alpha\beta}(z_i ) = g_{\alpha\beta}+a_i \gamma_\alpha \gamma_\beta\,$
with $\,a_i \equiv -(1/2+z_i )\,$ contains the off-shell parameter $z_i\,$~\cite{Pec69,Ols75}, where
$\,i=1,2,3\,$.
In the present work we show calculations for two parameter sets ``A" and ``B":
A) $\,a_i =0$ for all spin-3/2\ resonances,$\,$
B) $\,a_1 =0.12$, $a_2 =0.5$, $a_3 =0$ for the $D_{13}(1520)$,
and  $a_i =0$ for other resonances.
The fitted values of the couplings $(g_1,\, g_2)$ are: $\,$ (4.9, 5.27) for the $P_{33}(1232)$,
$\,$(7.25, 7.9) for the $D_{13}(1520)$ (on the proton), and (1.74, 4.75) for the $D_{33}(1700)$.

The cross section for real CS on the proton and
polarization observables are presented in~\figref{figu:2} for photon energies up to 1 GeV.
The cross section in the $\Delta$-resonance region agrees with the data as well as with
other calculations which are based on dispersion relations~\cite{Lvo97,Hun97,Dre00},
and an effective Lagrangian~\cite{Feu99}.
For comparison we show results of the calculations from Ref.~\cite{Lvo97} (dotted lines)
as they extend to high energies.
In Ref.~\cite{Lvo97} the imaginary part of the Compton amplitude is
calculated directly from pion-photoproduction data and the real part is
calculated using fixed-$t$ unsubtracted dispersion relations.
Our model and~\cite{Lvo97} give the same results for the photon asymmetry $\Sigma_\gamma$ at
the \Del\ resonance, however they predict a different slope.
Similar disagreement was noticed in~\cite{Feu99} --- it may point to the
importance of analiticity constraints in the calculation of the Compton amplitude.
The data~\cite{Ada93} for the photon asymmetry at $\theta_\gamma =90^\circ$  and  energy
$E_\gamma^{lab} \geq 500$ MeV are not
reproduced by the calculation with the set ``A". The description improves with the set ``B"
(dashed lines) though both parameter sets give very close results for cross sections (top panels).
For the photon angle 120$^\circ$ the difference between solid and dashed lines for $\Sigma_\gamma$
diminishes and the data do not allow to distinguish between the two parameter sets.
The recoil-proton polarization
$P_y^{pr}$ (\figref{figu:2}, bottom) vanishes below the pion-production threshold
since the imaginary part of the amplitude due
to $\pi N$ rescattering is zero~\footnote{Contribution to the
imaginary part from the $\gamma N$ rescattering is of the order $e^4$ and thus negligibly small}.
The behavior of $P_y^{pr}$ is in agreement with the present calculation as well as with the
model~\cite{Lvo97}, although the error bars on the data are large.

Above $\approx$ 900 MeV the present model loses predictive power as tails
of the higher resonances (not included in the model) will start to influence the cross section.

\section{Results of calculations and discussion} \label{sect:results}

\subsection{Cross sections for virtual Compton scattering} \label{subsec:results-cross}

Calculations of the exclusive differential cross sections for $\gamma p \to e^- e^+ p$
at two photon energies are shown in~\figref{figu:3}. The energy $E_\gamma^{lab}=320$ MeV
corresponds to the \Del \ resonance,
and $E_\gamma^{lab}=700$ MeV roughly corresponds to the region of the $D_{13}(1520)$.
At fixed energy the cross section depends on four essential
variables. In presenting the results we will therefore fix some of them.
The virtual-photon scattering angle $\theta_\gamma$ is chosen in the backward hemisphere
as our calculations indicate that this provides favorable conditions to suppress the BH contribution.

In~\figref{figu:3} the azimuthal lepton angle was fixed in plane, i.e.\ $\phi_l =0$ (coplanar kinematics).
For the case when $e^-$ and $e^+$ are not distinguished the angle combination
$\{\phi_l = 0,\,\theta_l\}$ is equivalent to $\{\phi_l = \pi,\, \pi -\theta_l\}$.
The results are shown as function of the polar angle $\theta_l$
at several values of the photon invariant mass.

The feature of the BH cross section (dashed lines) in coplanar kinematics is a pronounced peak
which develops
when either electron or positron moves along the direction of the initial photon.
The propagator of the lepton in the BH amplitude in~\figref{figu:1} (diagrams g, h)
becomes large in this case.
As is seen from~\figref{figu:3}, with increasing $M_\gamma$ the BH peak is shifted
towards larger angles  $\theta_l$, and for $M_\gamma= 500$ MeV (\figref{figu:3}, right, bottom)
it occurs at the angle $\,\theta_{BH}=32^\circ$.
The  maximal photon invariant mass at a given energy is determined from the
relation $\,M_\gamma^{max}=\sqrt{s}-m\,$ and corresponds to production of leptons
with energies $\epsilon_- = \epsilon_+ = \frac{1}{2}M_\gamma^{max}$ and momenta $\vec{k}_- = -\vec{k}_+$.
At this so-called kinematical limit the position of the peak reaches the maximal angle
$\;\theta_{BH}^{max} = \pi -\theta_\gamma$ which, for instance,
for $E_\gamma^{lab}=700$ MeV and $\theta_\gamma=135^\circ$
takes the value $45^\circ$ at $M_\gamma^{max}=543$ MeV.

An example of a calculation in non-coplanar kinematics is shown in~\figref{figu:4}.
It is seen that the BH peak disappears as the momentum of $e^-$ or  $e^+$
cannot be collinear to the photon momentum in this case. The $\phi_l$ dependence of the nuclear CS
cross section is governed by the interference RFs which stand in front of $\cos \phi_l$ and $\cos 2\phi_l$
in~\eqref{lt}. The $\phi_l$ dependence of the BH contribution is more complicated.

At small invariant masses the nuclear CS cross section (\figref{figu:3}, solid lines) shows a distinct
peaking at forward and backward angles, while at large invariant masses it tends to flatten.
This behavior can be understood from \eqref{ltcross}.
Assuming that $(M_\gamma / 2m_e )^2 \gg 1$ one can show that
the relative $e^-e^+$ momentum obeys the condition
\bea
|\vec{l}| \;\; &=&\;\;\;
 \Big\{ \;\; \begin{array}{cc}
\frac{1}{2}M_\gamma\; ( \sin^2\theta_l + M_\gamma^2/ \vec{k}^2 )^{-1/2} \,,
 & \text{if \hspace{0.1cm}  $M_\gamma \ll |\vec{k}|$ },\\
\frac{1}{2} M_\gamma\,,
& \text{ if \hspace{0.1cm} $|\vec{k}| \ll M_\gamma \sim    M_\gamma^{max} $ }.
\end{array}
\eqlab{l-condition}
\eea
Therefore at small $M_\gamma$ the phase-space factor $|\vec{l}|^3$ rapidly increases as
$\theta_l$ approaches  0$^\circ$ or 180$^\circ$.
The functions $W_T$ and $W_{TT}$ become the dominant contributions in~\eqref{lt}
and $\theta_l$ dependence of ${\cal S}$ is determined by $x_T$ and $x_{TT}$:
\bea
x_T = \frac{\sin^2 \theta_l +2 M_\gamma^2 / \vec{k}^2 }
{2(\sin^2 \theta_l + M_\gamma^2 / \vec{k}^2 ) }\,,
\;\;\;\;\;\;\;\;\;\;\;\;\;\;\;\;
x_{TT}= \frac{\sin^2 \theta_l }{2 (\sin^2 \theta_l + M_\gamma^2 / \vec{k}^2 ) }\,.
\eqlab{T-TT}
\eea
These are also rapidly changing functions, namely, $x_T$ increases from 1/2 to 1
and $x_{TT}$ decreases from 1/2 to 0 when $\theta_l$ approaches 0$^\circ$ or 180$^\circ$.
The product $|\vec{l}|^3 {\cal S}$ determines the behavior of the the cross
section shown in the top panels of~\figref{figu:3}.
At large invariant masses the phase-space factor is independent of the angle.
The $\theta_l$ dependence of the cross section then comes only from the factors $x_i$ in
Eqs.~(\ref{eq:xfactors}) and turns out to be rather flat for this particular kinematics.

Even if one moves away from the BH peak  there is still a BH background.
With increasing $M_\gamma$ its relative importance increases
as is seen from Figs.~(\ref{fig:figu:3},\ref{fig:figu:4}).
Therefore one has to choose a relatively large $\theta_\gamma$ to obtain the situation in
which the BH contribution can be ignored with respect to virtual CS.
There is a relation between $M_\gamma$ and the angle $\theta_\gamma^{min}$
which we define as the photon scattering angle, where
virtual CS is larger than BH by approximately
one order of  magnitude. The dependence is shown in~\tabref{table:2} for two energies
(for (almost) coplanar kinematics $\theta_l$ is
taken to be larger than  $\,\theta_{BH}\,$ to suppress the BH peak). As follows from~\tabref{table:2}
favorable conditions in the \Del\ region are at invariant masses up to $\approx 200$ MeV.
At larger values of $M_\gamma $ the BH background becomes comparable to or
larger than virtual CS at all angles $\theta_\gamma$.
At an energy of 700 MeV larger invariant masses up to $\approx$ 500 MeV become accessible.

\subsection{Results for response functions in virtual Compton scattering }
\label{subsec:results-response}

In Figs.~(\ref{fig:figu:5}-\ref{fig:figu:7}) the response $W_S=W_T +W_L$ and the ratios of RFs
(or elements of the density matrix defined in Eqs.~(\ref{eq:rho-w},\ref{eq:addit-LT})) are shown.
The calculations are presented with the parameter set ``B", as it gives better description
of the photon asymmetry in real CS.

First one notices from the angular distributions that at $\theta_\gamma = 0^\circ$ and 180$^\circ$
the interference  terms $\rho_{1-1}$ and $\rho_{10}$  vanish,
that can be explained from rotation symmetry around the beam
axis. This property serves as an additional test of the model.

In the \Del\ region (\figref{figu:5}) the transverse RF is dominant over the whole interval of $M_\gamma$.
We studied the contribution of the additional coupling $G_3(k^2)$ in the $\gamma N \Delta$ vertex
in~\eqref{vdelta} by varying $g_3$ within the
limits $\pm g_2$. The effect turns out to be extremely small.
This is due to small $k^2 \le 0.08$ GeV$^2$ which can be reached at this energy, and the dominance
of the magnetic $f^{1+}_{MM}$ transition, which is very weakly dependent on $G_3(k^2)$.
There is an effect of $\sigma$ exchange on the interference
RFs, or on $\rho_{1-1}$ and $\rho_{10}$ (compare solid and dashed curves in~\figref{figu:5}).
This is interesting in view of the very small
contribution of sigma exchange to the cross section near the \Del\ peak.
However, the region of $M_\gamma$ above $\approx$ 150 MeV will be
difficult to access due to large BH contribution (see~\tabref{table:2}).
The angular distribution of $W_S$ at fixed $M_\gamma$ is similar to real CS cross section
$d\sigma (\gamma)/d\Omega_\gamma$, and is determined by the dominant $M1$ multipole
and its interference with $E1$ and $E2$ multipoles.

At a higher photon energy, shown in Figs.~(\ref{fig:figu:6},\ref{fig:figu:7}),
the transverse response is still rather substantial.
In Fig.~(\ref{fig:figu:6}) the effects of the important contributions
are shown by switching off the corresponding process.
The effect of the $D_{13}(1520)$ resonance is clearly seen in a pronounced angular distribution
while other contributions give rise to a flat background (dotted lines)
with $\Im \rho_{10}$ as an exception.  This background in  $\rho_{1-1}$ and $\Re \rho_{10}$
is also independent of $M_\gamma$ as is seen in the left panel.
Other important contributions in this energy region are the
$S_{11}(1535)$ excitation and the $\sigma$ exchange.
The $S_{11}$ resonance contributes 10-20\% to the cross section at $M_\gamma = 300$ MeV,
and also to $\rho_{00}$ and $\Re \rho_{10}$ at backward angles; its role increases
with increasing invariant mass.
$\sigma$ exchange, as a t-channel contribution, shows up at backward angles
with the exception of $\Im \rho_{10}$, where it is seen at forward angles.
There is also a small contribution to $\rho_{1-1}$ from the $D_{33}(1700)$,
however, its effect is not shown separately because the e.m.\ couplings of this resonance
could not be fixed quite accurately from pion photoproduction.
We should emphasize that all contributions add coherently in the total amplitude
resulting in a strong interference.
In particular, the $P_{33}(1232)$ resonance, through an interference,
affects observables in the 2nd resonance region.

For comparison we also present calculations with the parameter set ``A" (long-dashed lines).
In $W_S$ large differences between the set ``A" and set ``B" appear at $\theta_\gamma < 90^\circ$
and large invariant masses, on average the differences are about 20\%.
The most striking effects can be observed in
$\rho_{00}$, $\rho_{1-1}$ and $\Re \rho_{10}$ (compare solid and long-dashed lines in~\figref{figu:6}).
The big difference in $\rho_{1-1}$ at 90$^\circ$ is similar to that for the photon asymmetry
in~\figref{figu:2}, while at larger angles the difference is reduced.
As is seen from Fig.~(\ref{fig:figu:6}) for $\rho_{00}$ and $\Re \rho_{10}$,
the longitudinal response also proves to be rather sensitive to
parameters $a_{1,2}$ for the $D_{13}$.

In principle one might expect a contribution of the Roper resonance,
$P_{11}(1440)$, to the longitudinal response. Data on pion electroproduction
(see review~\cite{Bur96}) indicate however that the corresponding
helicity amplitude $S_{1/2}$ is very small, consistent with zero. The longitudinal
coupling in the $\gamma N P_{11}$ vertex has thus been neglected. The
transverse coupling is known to be small from real-photon data.

In~\figref{figu:7} the effect of $G_{3}(k^2)$ in the $\gamma N D_{13}(1520)$ vertex is demonstrated.
The dashed curves are calculated with $g_3 = -3.3$ which is chosen to reproduce the
$k^2$ dependence of the ratio of the helicity amplitudes $A_{1/2}$ and $A_{3/2}$
(or the corresponding multipoles $E_{2-}^{1/2}$ and
$M_{2-}^{1/2}$) for pion electroproduction~\cite{Bur96}.
As is seen, the $G_3(k^2)$ coupling has a considerable effect on $W_S$ and thus on the cross section.
The reason is that $G_3(k^2)$ strongly influences the electric dipole transition $E1$
and the latter in turn becomes the dominant term at large positive $k^2$.
For the following consideration we choose $\sqrt{s} = M_R$ ($M_R$ is a resonance mass),
then the 3-momentum
squared of the virtual photon is
\bea
\vec{k}^2 = k_0^2 - k^2 =
\frac{1}{4M_R^2}[\,(M_R+m)^2-k^2\,]\,[\,(M_R-m)^2-k^2\,]\,.
\eqlab{3-momentum}
\eea
It is seen that $\vec{k}^2$ decreases in the time-like region and vanishes
at the kinematical limit, where $k_0 = M_\gamma^{max} = M_R -m$.

The behavior of RFs for $e^-e^+$ production can be understood
based on general properties of the resonance multipoles at small $|\vec{k}| \ll R^{-1}$, where
$R$ is a typical interaction radius. According to~\cite{Eis70} (ch. 6.2 and 6.3)
one has for the e.m. transition with multipolarity $j$
\bea
T(\vec{k};Mj) \propto |\vec{k}|^j\,,\;\;\;\;\;\;\;\;\;\;
T(\vec{k};Ej) \propto |\vec{k}|^{j-1} k_0\,,\;\;\;\;\;\;\;\;\;\;
T(\vec{k};Lj) \simeq - T(\vec{k};Ej)\, \Big( \frac{j}{j+1} \Big)^{1/2}\,.
\eqlab{multipoles}
\eea
Therefore for the $D_{13}$ resonance magnetic quadrupole ($M2$) vanishes,
and electric dipole ($E1$) and longitudinal
dipole ($L1$) remain finite at $k^2 = (M_\gamma^{max})^2$.
At the same time all multipoles for the $P_{33}(1232)$ resonance, magnetic dipole ($M1$),
electric and longitudinal
quadrupoles ($E2$ and $L2$), are proportional to $|\vec{k}|$ and go to zero.
To have a more quantitative estimate we can use predictions of the non-relativistic
quark model (e.g., Ref.~\cite{Bij96}). In~\tabref{table:3} expressions are collected
for the e.m. helicity amplitudes and the resonance pion-production multipoles.
It follows from~\tabref{table:3} that:

i) $A_{1/2}$ and $A_{3/2}$ for the $D_{13}$ are finite at small $|\vec{k}| $,
while those for the $P_{33}$ vanish;

ii) if $|\vec{k}| \to 0$ the  ratio $A_{1/2} / A_{3/2}$ for the $D_{13}$ takes the value $1/\sqrt{3}$
which coincides with the corresponding ratio for the $P_{33}$ (in this and other $SU(6)$ symmetrical models);

iii) as $k^2$ increases towards the maximal value
the spin-flip contribution (terms $\propto \vec{k}^2$) diminishes for the $D_{13}$, correspondingly
$E_{2-}$ increases and $M_{2-}$ decreases (absolute values).

Since the $E1$ transition dominates the $D_{13} \to \gamma^* N$ process,
the coupling $G_3(k^2)$ affects strongly the cross section in~\figref{figu:7}.
The transverse  response $\,W_T$ at the resonance position is roughly proportional
to $\,|E1|^4$, while the longitudinal response $\,W_L$ to $\,|L1|^2\,|E1|^2$.
At small $|\vec{k}|$ one can make use of Siegert's theorem
(last relation in Eqs.~(\ref{eq:multipoles})) which relates $L1$ and $E1$ amplitudes.
From this it can be seen that the ratio $\,\rho_{00}=W_L / (W_T +W_L)\,$
is equal to $1/3$,  irrespective of $G_3(k^2)$.
Other elements of the density matrix on~\figref{figu:7}, except  $\Im \rho_{10}$, show a dependence
on $G_3(k^2)$.

It is of interest to compare Figs.~(\ref{fig:figu:6},\ref{fig:figu:7}) for the $D_{13}$
with~\figref{figu:5} for the $P_{33}$ resonance.
Due to a decrease of the $M1$ intensity as a function of $M_\gamma$
the response $W_S$ for the $\Delta$ resonance falls off towards the maximal $M_\gamma$.
The longitudinal response $W_L$ contains the small resonance multipole
$L2$ which also vanishes at $ |\vec{k}| \to 0 $, while the background receives
nonvanishing contributions from $L1$ multipoles
(such as $L_{0+}$ and $L_{2-}$ in pion electroproduction).
These cause an increase of $\rho_{00}$ at large photon masses in~\figref{figu:5}.
This effect comes as a result of a balance between $G_1(k^2)$ and $G_2(k^2)$ contributions
in the $\gamma N \Delta$ vertex in~\eqref{vdelta}.
Note also that in the quark model in~\tabref{table:3} the electric quadrupole is absent, $E2 =L2 =0$,
which corresponds to
a particular choice of the couplings $\,2m g_1 = M_{\Delta} g_2 = M_{\Delta} g_3$.

The features of the e.m decay of the resonances in the time-like region are, in general,
different from those studied in electron scattering, where
$|\vec{k}|$ is always larger than $k_0$.
For instance, in pion photo- and electroproduction  interesting properties of the
$D_{13}(1520)$ resonance have been observed
at $-3\,$GeV$^2\,<\, k^2\,\leq \,0$ (see, e.g.~\cite{Bur96}). In terms of the multipoles,
the ratio $E1/M2 \approx -1$ at large negative $k^2$, crosses zero at $k^2 \approx -1$ GeV$^2$
and is about +2.1 at the real-photon point.
It would be of interest to see if this ratio keeps on increasing at positive $k^2$
and becomes very large near $k^2 = (M_{D_{13}} - m)^2 = 0.3364$ GeV$^2$.

Finally we note that the increase of $W_S$  in~Figs.~(\ref{fig:figu:6},\ref{fig:figu:7})
above $M_\gamma \approx 500$ MeV is due to the fact that the e.m.\
form factors contain the $\rho$-meson propagator.
As $M_\gamma$ approaches the $\rho$-meson pole (which would be possible at photon energies
above 1 GeV), the process $\gamma p \to e^- e^+ p\,$ proceeds through creation
of the $\rho$ meson, as assumed in VMD models.

\section{Conclusions} \label{sect:conclusions}

We investigated virtual Compton scattering on the proton in the time-like region
($\gamma p \to e^- e^+ p\,$). When in an experiment the $e^-$ and $e^+$ are not distinguished,
the Bethe-Heitler-nuclear interference vanishes and the cross
section is the incoherent sum of cross sections for the BH and the nuclear processes.
We have shown that, contrary to common
preconception, a considerable part of the phase space, mainly at backward angles for the
virtual photon, is hardly contaminated by
the Bethe-Heitler process. Under these  conditions it is possible to decompose
the exclusive cross section in terms of response functions.
These are directly related to the photon density matrix  $\rho_{\lambda \lambda'}$
which characterizes polarization properties of the virtual photon.
This offers a possibility to analyze $e^- e^+$ production experiments also in terms of matrix
elements $\rho_{\lambda \lambda'}$ or, equivalently,
tensor and vector polarizations of the time-like photon.

Differential cross sections and matrix elements  $\rho_{\lambda \lambda'}$ are
calculated in a unitary
K-matrix model which includes nucleon, mesons, and baryon resonances with masses up to 1.7 GeV.
The model is tested for real-photon Compton scattering at energies up to 1 GeV.
The agreement with data is reasonable and of a comparable quality to
recent K-matrix calculation~\cite{Feu99} and dispersion-relation analysis~\cite{Lvo97}.

The dilepton production  in the $\Delta$-resonance region is
dominated by the transverse response. At the largest $M_\gamma$ at this energy
certain elements $\rho_{\lambda \lambda'}$ show the effect of $\sigma$ exchange.
The latter could be an approximation to various t-channel scalar-isoscalar exchanges
(e.g., two-pion, $\epsilon(760)$ and $f_0$(400--900)).

Photon energies corresponding to the 2nd resonance region, of about 700 MeV,
allow one to explore higher photon masses up to $\approx$ 500 MeV.
Main mechanisms contributing here are the $D_{13}(1520)$-resonance excitation and, to a lesser extent,
the $S_{11}(1535)$ resonance and $\sigma$ exchange.
The transverse-transverse element $\rho_{1-1}$ strongly depends on the so-called off-shell
parameters $a_1$ and $a_2$ in the e.m.\ vertex of the $D_{13}$.
The longitudinal  $\rho_{00}$ and longitudinal-transverse $\Re \rho_{10}$ elements
turn out to be sensitive to all contributions and
with increasing photon invariant mass this sensitivity becomes more pronounced.
At large $M_\gamma$ the cross section and the density matrix become strongly
dependent on the part of the e.m. vertex of the $D_{13}(1520)$
which contributes only for virtual photons.

Response functions are  thus shown to be
an important tool to distinguish between different mechanisms in the $e^- e^+$ production
in the resonance region. If the resonance contributions can be separated, this will allow for
studying their e.m. properties in the time-like region which may give
an information supplementary to that obtained from the electron scattering.

\section*{Acknowledgments}
This work is supported by the Foundation for Fundamental Research of the Netherlands (NWO).
A.\ Yu.\ K.\ acknowledges a special grant from the NWO.
The authors thank  A.\ I.\ L'vov for sending results of calculations in the
dispersion-relation approach. We thank also J.\ Bacelar for stimulating
discussions on the experimental feasibility of virtual Compton scattering
and valuable suggestions.
Discussions with J.\ Messchendorp, M.\ Mostovoy and R.\ Timmermans are highly appreciated.

\appendix

\section{Response functions in helicity formalism}

It is convenient to introduce the following set of polarization vectors for a time-like photon
\bea
\epsilon^*(0)=\frac{1}{M_\gamma}(|\vec{k}|,0,0,k_0 )\,,\;\;\;\;\;\;
\epsilon^*(\pm 1)=\frac{1}{\sqrt{2}}(0,\mp 1,i,0 )
\eea
which satisfy  $\epsilon^* (\lambda)\cdot k =0$ and
\bea
\epsilon^*(\lambda)\cdot \epsilon(\lambda') =-\delta_{\lambda \lambda'}\,,\;\;\;
\sum_{\lambda=0,\pm 1} \epsilon^{*\mu}(\lambda) \epsilon^\nu(\lambda)
=-g^{\mu\nu}+\frac{k^\mu k^\nu}{M_\gamma^2}\,.
\eea

The RFs defined in \eqref{response} can now be written as
\bea
W_T =\frac{1}{4}\sum_{polar.}&& |J\cdot\epsilon^*(+1) |^2+|J\cdot\epsilon^*(-1) |^2=
\frac{1}{4}\sum_{\Lambda ,\Lambda'=\pm 1/2,\,\lambda=\pm 1 } |f_{+1\Lambda',\lambda\Lambda}|^2
+|f_{-1\Lambda',\lambda\Lambda}|^2\,,
\nonumber\\
W_L =\frac{1}{4}\sum_{polar.}&& |J\cdot\epsilon^*(0) |^2=
\frac{1}{4}\sum_{\Lambda ,\Lambda'=\pm 1/2,\,\lambda=\pm 1 }|f_{0 \Lambda',\lambda\Lambda}|^2\,,
\nonumber\\
W_{TT} =\frac{1}{4}\sum_{polar.} && 2\Re\, J\cdot\epsilon^*(+1) [J\cdot\epsilon^*(-1) ]^* =
\frac{1}{4}\sum_{\Lambda ,\Lambda'=\pm 1/2 ,\,\lambda=\pm 1}
 2 \Re\, f_{+1\Lambda',\lambda\Lambda}f^*_{-1\Lambda',\lambda\Lambda}\,,
\nonumber\\
W_{LT} =\frac{1}{4}\sum_{polar.}  && 2\Re\, J\cdot\epsilon^*(0) [J\cdot\epsilon^*(+1) -
J\cdot\epsilon^*(-1)]^*  \nonumber\\
&=&\frac{1}{4}\sum_{\Lambda ,\Lambda'=\pm 1/2 ,\,\lambda=\pm 1} 2\Re\,
f_{0\Lambda',\lambda\Lambda}^* [ f_{+1\Lambda',\lambda\Lambda}-f_{-1\Lambda',\lambda\Lambda}] \,,
\eqlab{Aresponse}
\eea
where we introduced the helicity amplitude
\bea
f_{\lambda'\Lambda',\lambda\Lambda} =e\epsilon^*_\mu(\lambda')
\bar{u}(p' ,\Lambda') M^{\mu\nu}u(p ,\Lambda) \epsilon_\nu(\lambda) \;
\eea
which is a function of $s, \theta_\gamma$ and $M_\gamma$.
The additional RF $\tilde{W}_{LT}$, which defines the vector polarization
$p_y=i\frac{2}{\sqrt{3}} t_{11}$ of the virtual photon (see \eqref{t-w}),
can be calculated using the last formula in \eqref{Aresponse}, where the operation $\Re\, ...$ should be
replaced by  $\Im\, ...$.

Due to space-reflection invariance there are 8 (4) independent transverse
(longitudinal) amplitudes in the sums in Eqs.~(\ref{eq:Aresponse}).
They can be chosen according to~\tabref{table:4}. The remaining amplitudes are determined
from the relation
$f_{-\lambda' -\Lambda', -\lambda -\Lambda} = (-1)^{\lambda' -\Lambda' -(\lambda-\Lambda )}
f_{\lambda' \Lambda', \lambda \Lambda}$~\cite{Jac59}.

In terms of $f_i$ of~\tabref{table:4} one can rewrite the response functions as
\bea
W_T &=& \frac{1}{2}\sum_{i=1}^8 |f_i|^2\,,\;\;\;\;\;\;\;\;\;\;\;\;\;\;\;
W_L = \frac{1}{2}\sum_{i=9}^{12} |f_i|^2\,,\nonumber\\
W_{TT}&=& \Re\, (\,f_1 f_3^* +f_2 f_4^* +f_5 f_7^* + f_6 f_8^* \,)\,,\nonumber\\
W_{LT}&=& \Re\, [\, f_9^* (f_1-f_3 ) -f_{10}^* (f_2 -f_4 ) +f_{11}^* (f_5-f_7 ) - f_{12}^* ( f_6 -f_8 )\, ]\,.
\eea

For real photons due to the time reversal~\cite{Jac59} one has
in addition $f^r_7 =f^r_3$ and $f^r_8 =-f^r_4$,
where the superscript ``$r$" indicates the real CS limit:
$f^r_{\lambda'\Lambda', \lambda\Lambda}(s,\theta_\gamma ) =
f_{\lambda'\Lambda', \lambda\Lambda}(s,\theta_\gamma,M_\gamma =0)$.
In this case
\bea
&&W^r_T =\frac{1}{2}[\,|f^r_1 |^2 +|f^r_2 |^2 + 2|f^r_3 |^2 + 2|f^r_4 |^2 + |f^r_5 |^2 +
|f^r_6 |^2\,]\,,\nonumber\\
&&W^r_{TT}= \Re \,[\,f_3^{r*} ( f^r_1 +f^r_5 ) + f_4^{r*} ( f^r_2 - f^r_6 )\, ]
\eqlab{Arealcs}
\eea
since $f_9^r, \cdots, f_{12}^r$ vanish.
\eqref{Arealcs} agrees with definitions given in~\cite{Pfe74,Ish80} up to a common normalization factor.

In addition to the photon asymmetry in~\eqref{asymmetry} we will also calculate
the polarization of the recoil proton in real CS.
It can be expressed through the helicity amplitudes as
\bea
P_y^{pr} = \frac{1}{4 W^r_{T}} \sum_{\Lambda=\pm 1/2 ,\,\lambda, \lambda' =\pm 1} && 2\Im\,
f^r_{\lambda' -1/2,\lambda \Lambda} f_{\lambda' +1/2,\lambda \Lambda }^{r*}  \nonumber\\
&=& -\frac{1}{W^r_T } \Im\, [\, f_4^{r*}(f^r_1 +f^r_5 )+f_3^{r*} (f^r_6 -f^r_2 ) \,]\,.
\eea



\begin{table}
\begin{tabular}{cc|cc}
  \multicolumn{2}{c|}{$E_\gamma^{lab} = 320$ MeV \,($M_{\gamma}^{max} = 278$ MeV)}
& \multicolumn{2}{c}{$E_\gamma^{lab} = 700$ MeV \,($M_{\gamma}^{max} = 543$ MeV) }     \\
  $M_\gamma$ (MeV) &  $\theta_\gamma^{min}$ (deg) &  $M_\gamma$ (MeV)  & $\theta_\gamma^{min}$ (deg) \\
\hline
    5   & 30  & 5 & 30   \\
    50  & 90  & 50 & 60  \\
    100 & 120 & 100 & 70 \\
    150 & 150 & 150 & 90 \\
    200 & 170 & 250 & 110 \\
        &     & 350 & 130 \\
        &     & 450 & 150 \\
        &     & 530 & 170 \\
\end{tabular}
\caption{ Minimal photon angle (defined in sect.~\protect\ref{subsec:results-cross})
as function of photon invariant mass. }
\tablab{table:2}
\end{table}

\begin{table}
\begin{tabular}{c|cc}
                &  $D_{13}(1520),\;\;$I=1/2$\;$ (proton) &  $P_{33}(1232),\;\;$I=3/2              \\
\hline
\\
$(\frac{k_0}{\pi})^{1/2} A_{1/2}$  &   $2i (\vec{k}^2-b^2) a F(\vec{k}^2)\mu_q $   &
                                           $- \frac{2\sqrt{2}}{3} |\vec{k}| F(\vec{k}^2)\mu_q$ \\
$(\frac{k_0}{\pi})^{1/2} A_{3/2}$   &   $-2i \sqrt{3}  b^2 a F(\vec{k}^2)\mu_q$      &
                                     $- \frac{2\sqrt{2}}{\sqrt{3}} |\vec{k}| F(\vec{k}^2)\mu_q$ \\
${A_{1/2}}/{A_{3/2}}$ & $\frac{1}{\sqrt{3}} (1-\frac{\vec{k}^2}{b^2})$ & $\frac{1}{\sqrt{3}}$ \\
$E_{2-}^I\,\;(E1)$  & $\frac{i}{\sqrt{3}} (\vec{k}^2-4b^2) a F(\vec{k}^2)\mu_q f$ & $-$  \\
$E_{1+}^I\,\;(E2)$  & $-$ & $ 0 $                                                 \\
$M_{1+}^I\,\;(M1)$   & $-$ & $\frac{4}{\sqrt{3}} |\vec{k}| F(\vec{k}^2) \mu_q f$  \\
$M_{2-}^I\,\;(M2)$  & $-\frac{i}{\sqrt{3}}\vec{k}^2 a F(\vec{k}^2) \mu_q f$ & $-$  \\
\end{tabular}
\caption{ Electromagnetic helicity amplitudes and pion-production multipoles
          for $D_{13}(1520)$ and $P_{33}(1232)$ resonances
          in model~\protect\cite{Bij96}.\,~\protect$F(\vec{k}^2) = (1+a^2 \vec{k}^2)^{-2} ,
          \,b^2 = m_q k_0/g_q$, and $m_q, \mu_q, g_q$ and $a$ are respectively
          constituent quark mass, scale magnetic moment, gyromagnetic ratio, and scale parameter.
The constant~\protect$f = ( \pi / k_0 )^{1/2}
( k_\gamma m \Gamma_{\pi N} / 4 \pi q_\pi M_R \Gamma_{tot}^2  )^{1/2}$
relates helicity amplitudes to e.m. multipoles, where~\protect$k_\gamma$ is equivalent
real-photon energy,~\protect$q_\pi$ is pion 3-momentum,~\protect$\Gamma_{\pi N}\,(\Gamma_{tot})$
is one-pion (total) decay width of the resonance. }
\tablab{table:3}
\end{table}

\begin{table}
\begin{tabular}{cc|cc|cc}
$f_1$ & $f_{+1\,+1/2,\,+1\,+1/2}$ & $f_5$ & $f_{+1\,-1/2,\,+1\,-1/2}$ & $f_{9}$  & $f_{0\,+1/2,\,+1\,+1/2}$\\
$f_2$ & $f_{-1\,-1/2,\,+1\,+1/2}$ & $f_6$ & $f_{-1\,+1/2,\,+1\,-1/2}$ & $f_{10}$ & $f_{0\,-1/2,\,+1\,+1/2}$\\
$f_3$ & $f_{-1\,+1/2,\,+1\,+1/2}$ & $f_7$ & $f_{-1\,-1/2,\,+1\,-1/2}$ & $f_{11}$ & $f_{0\,-1/2,\,+1\,-1/2}$\\
$f_4$ & $f_{+1\,-1/2,\,+1\,+1/2}$ & $f_8$ & $f_{+1\,+1/2,\,+1\,-1/2}$ & $f_{12}$ & $f_{0\,+1/2,\,+1\,-1/2}$\\
\end{tabular}
\caption{Independent helicity amplitudes in the $\gamma p\rightarrow e^- e^+ p$ reaction.}
\tablab{table:4}
\end{table}


\begin{figure}[hbt]
\begin{center}
\leavevmode
\epsfysize=15cm
\rotate[r]
{\epsfbox[100 81 357 775]{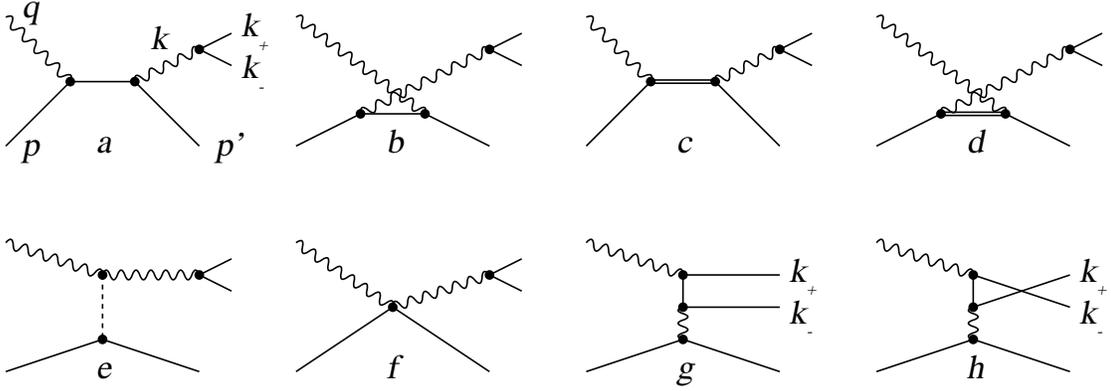}}
\caption{Tree-level graphs describing the $\gamma N \to e^- e^+ N$ process.
Diagrams a -- f correspond to virtual CS on the nucleon, g and h  to the Bethe-Heitler contribution.
Single (double) solid line depicts a nucleon (baryon resonance), wavy line a photon,
dashed line  $\pi^0,\,\sigma$ and $\eta$ mesons.
Diagram f describes a possible contact $\gamma \gamma^* N N$ vertex.}
\figlab{figu:1}
\end{center}
\end{figure}

\begin{figure}[hbt]
\begin{center}
\leavevmode
\epsfysize=15cm
{\epsfbox[51 139 488 549]{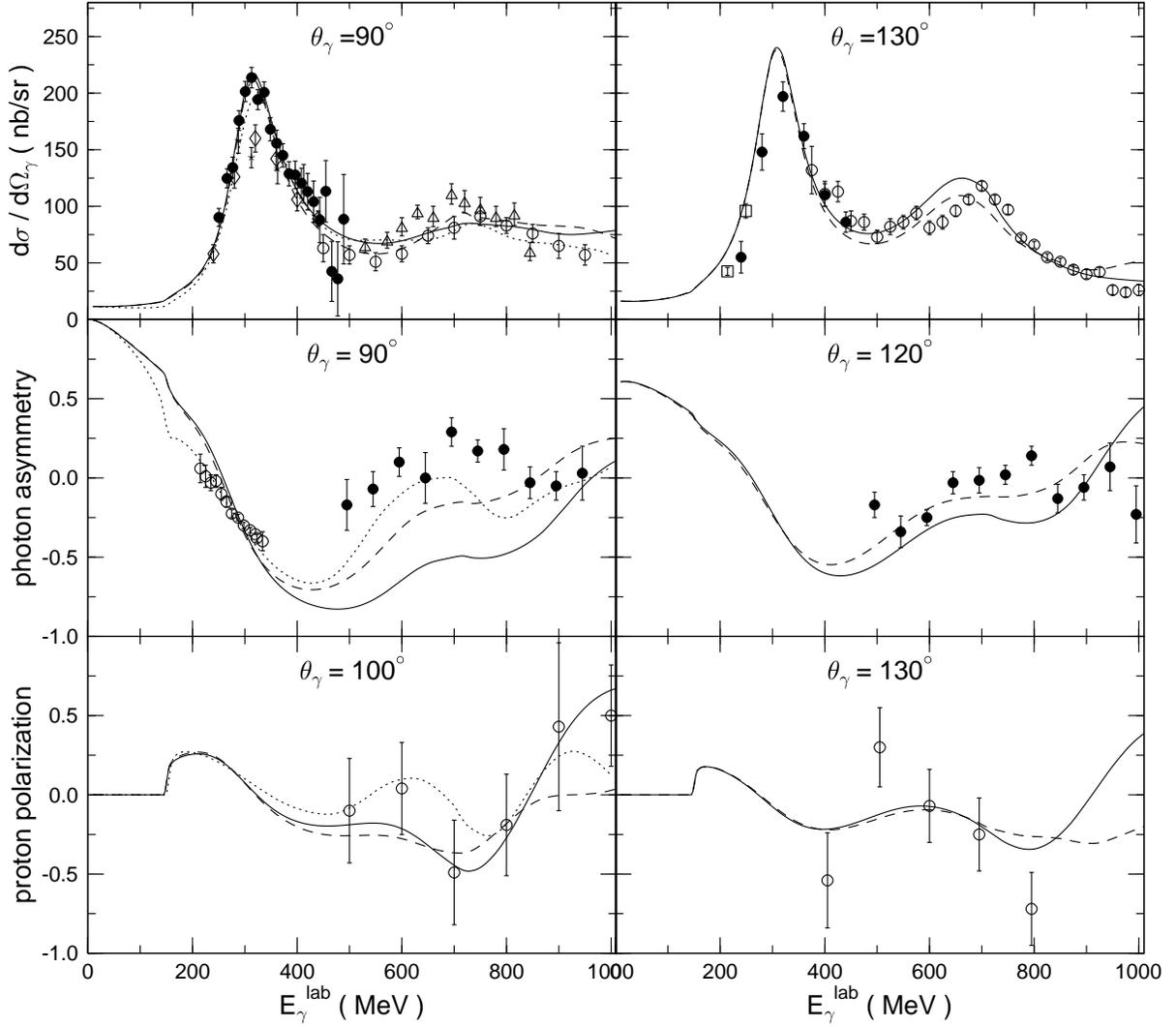}}
\caption{Differential cross section (top), photon asymmetry (middle), and
recoil-proton polarization (bottom) for real-photon Compton scattering as function of photon energy.
Solid (dashed) lines are the present model calculations with the parameter set ``A"\,(``B"),
dotted lines are calculations from~\protect\cite{Lvo97}.
Data for cross sections are: left panel - $\bullet$ Mainz~\protect\cite{Hun97},
$\star$ Cornell~\protect\cite{Dew61},
$\triangle$ MIT~\protect\cite{Sti63},
$\diamond$ Bonn~\protect\cite{Gen76},
$\circ$ Tokyo~\protect\cite{Tos78};\,
right panel - $\bullet$ Bonn~\protect\cite{Gen76}, $\circ$ Tokyo~\protect\cite{Ish80},
$\Box$ Moscow (Lebedev Inst.)~\protect\cite{Bar66}.
Data for photon asymmetry: $\circ$ LEGS Collaboration~\protect\cite{Bla96},
$\bullet$ Erevan~\protect\cite{Ada93};\,
and for proton polarization: $\circ$ Tokyo~\protect\cite{Wad84}.         }
\figlab{figu:2}
\end{center}
\end{figure}

\begin{figure}[hbt]
\begin{center}
\leavevmode
\epsfysize=15cm
{\epsfbox[44 88 561 772]{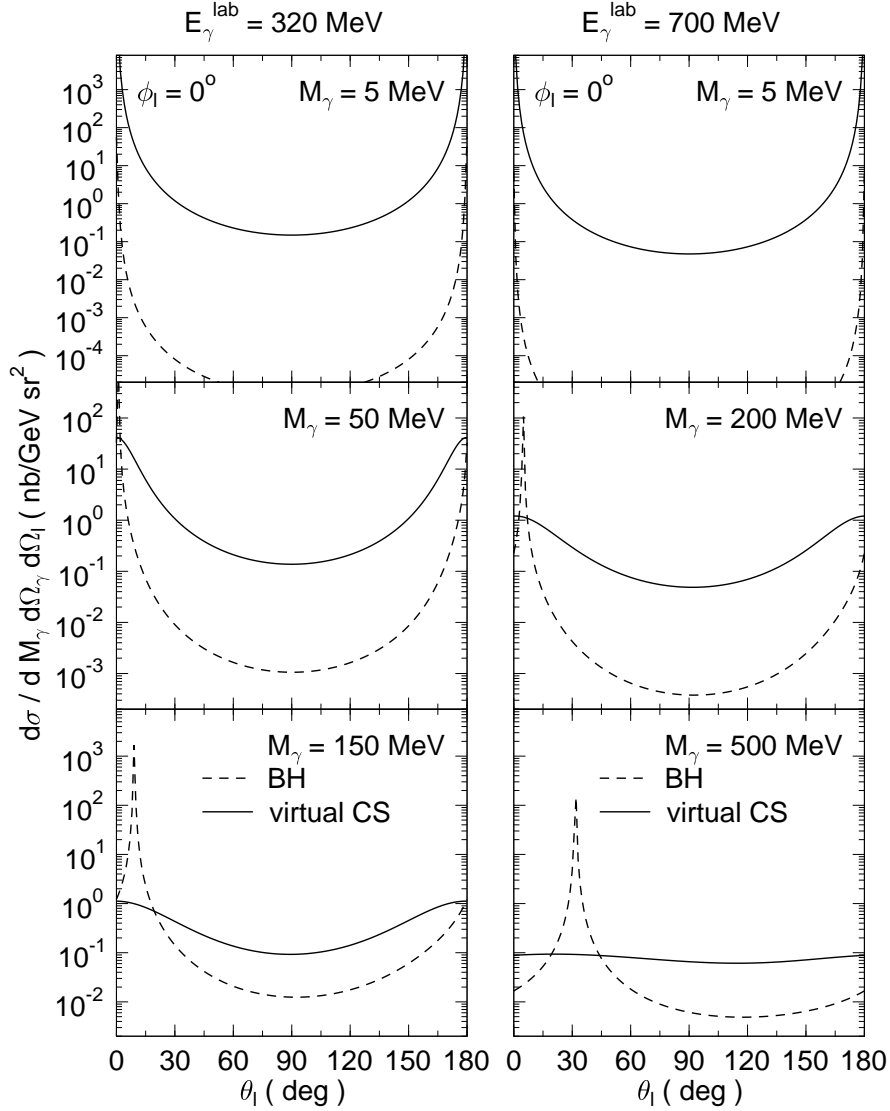}}
\caption{Exclusive differential cross section in the c.m. as function of
lepton ``sharing'' angle $\theta_l$ in coplanar kinematics.
Left panel corresponds to $E_\gamma^{lab} = 320$ MeV, right panel to $E_\gamma^{lab} = 700$ MeV.
The virtual-photon angle is $\theta_\gamma = 135^\circ$.
Dashed and solid lines are respectively the Bethe-Heitler
and the virtual Compton scattering (parameter set ``A") contributions.   }
\figlab{figu:3}
\end{center}
\end{figure}

\begin{figure}[hbt]
\begin{center}
\leavevmode
\epsfysize=10cm
{\epsfbox[44 500 322 743]{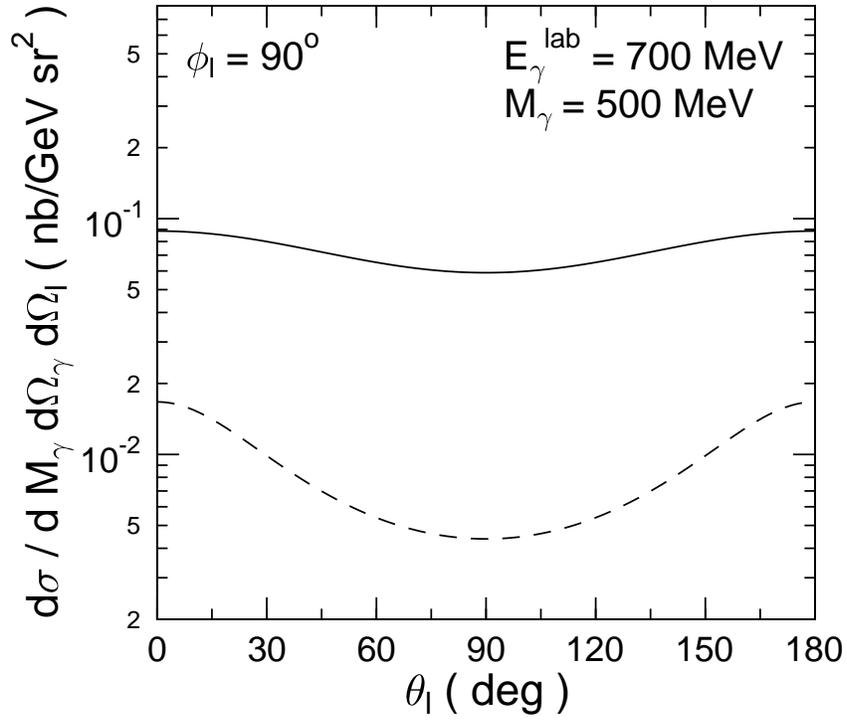}}
\caption[f3]{Exclusive cross section at an incoming energy of 700 MeV in non-coplanar kinematics.
Notation for the curves is the same as in~\protect\figref{figu:3}.}
\figlab{figu:4}
\end{center}
\end{figure}

\begin{figure}[hbt]
\begin{center}
\leavevmode
\epsfysize=15cm
{\epsfbox[63 232 452 776]{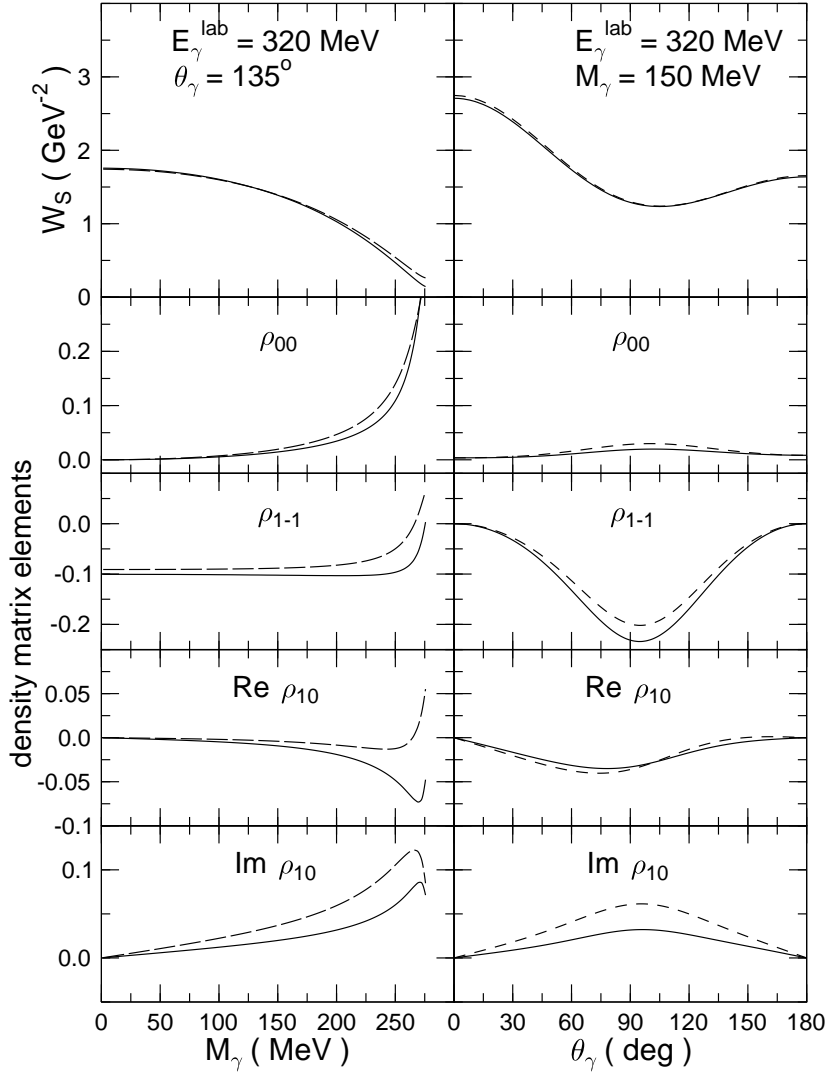}}
\caption{Response $W_S =W_T + W_L$ (top) and
polarization density matrix  versus virtual-photon invariant mass
$M_{\gamma}$ (left) and photon angle $\theta_\gamma$ (right).
Incoming photon energy is $320$ MeV.
Solid lines are calculated with the total amplitude, dashed lines without $\sigma$ exchange. }
\figlab{figu:5}
\end{center}
\end{figure}

\begin{figure}[hbt]
\begin{center}
\leavevmode
\epsfysize=15cm
{\epsfbox[63 232 452 776]{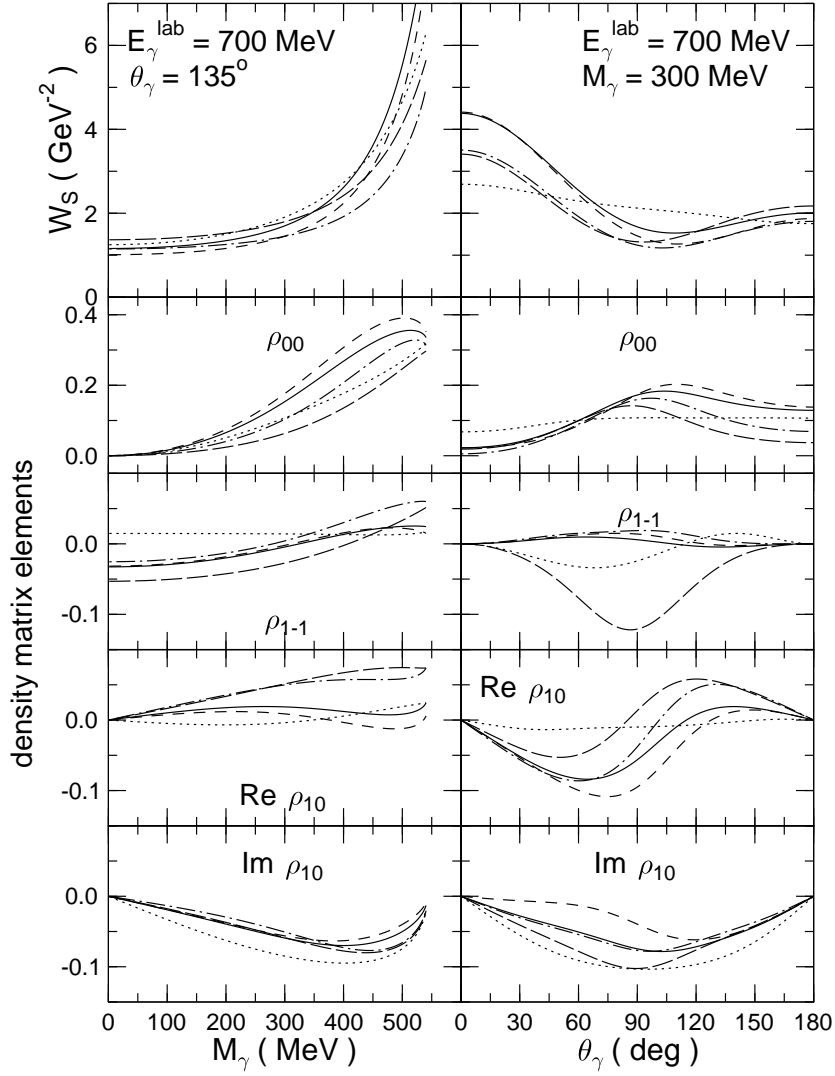}}
\caption{The same as in~\protect\figref{figu:5} but at photon energy $700$ MeV.
Solid lines show the calculation with the total amplitude (parameter set ``B"),
dotted lines without $D_{13}(1520)$ resonance,
dash-dotted lines without $S_{11}(1535)$ contribution,
short-dashed lines without $\sigma$ exchange,
and long-dashed lines are calculated with the total amplitude (parameter set ``A"). }
\figlab{figu:6}
\end{center}
\end{figure}

\begin{figure}[hbt]
\begin{center}
\leavevmode
\epsfysize=15cm
{\epsfbox[57 240 278 767]{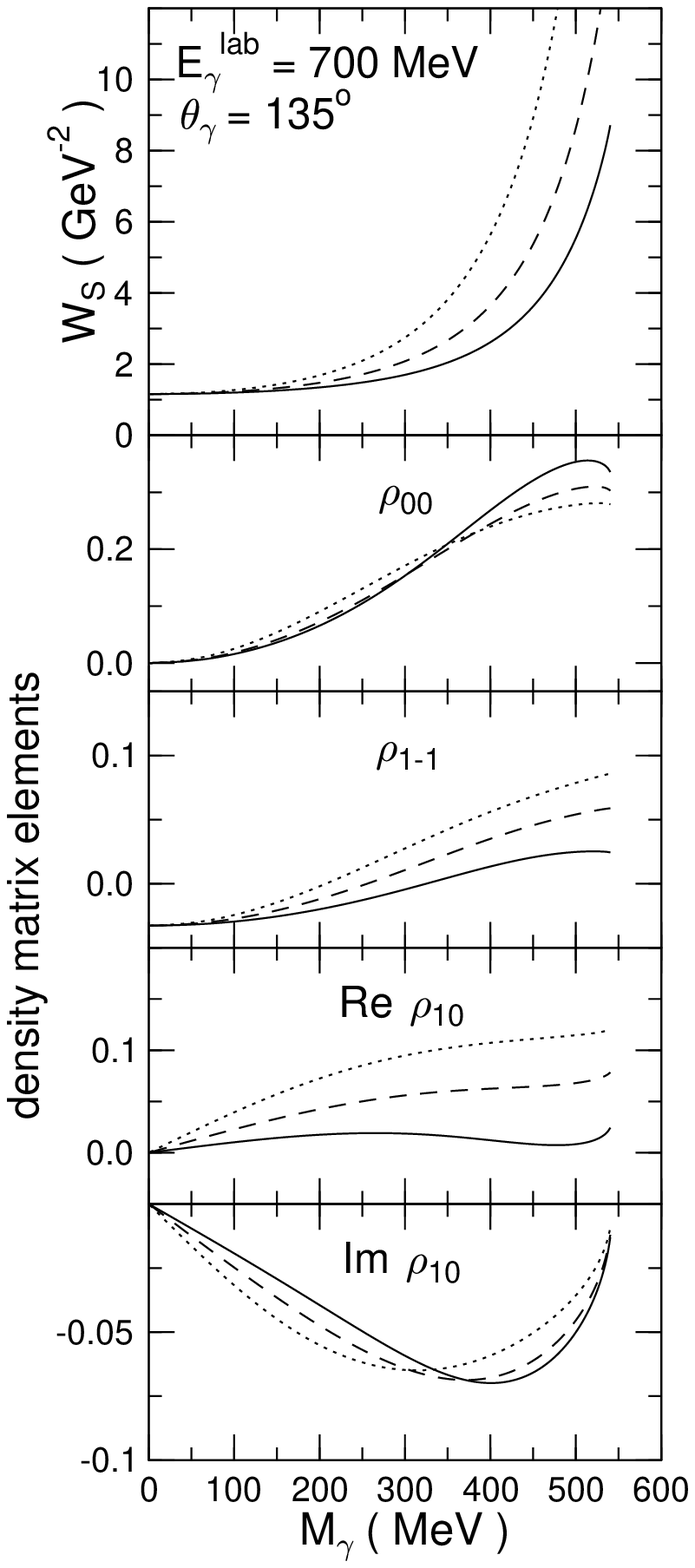}}
\caption{Response $W_S =W_T + W_L$ (top) and $\rho_{\lambda \lambda'}$
calculated for different values of $g_3$ in the $\gamma N D_{13}$ vertex~\protect\eqref{vdelta}.
Solid lines:\, $g_3  =0$, dashed lines:\, $g_3 = -3.3$ (from pion electroproduction),
and dotted lines:\, $g_3 = -7.9$. Couplings $g_1$ and $g_2$ are respectively 7.25 and 7.9. }
\figlab{figu:7}
\end{center}
\end{figure}


\begin{thebibliography}{References}

\bibitem{Procee} Proceedings of Workshop on Virtual Compton Scattering, Clermont-Ferrand, France, 1996.

\bibitem{Gui98} P. A. M. Guichon and M. Vanderhaeghen, Progr. Part. Nucl. Phys. {\bf 41}, 125 (1998).

\bibitem{Sch95} M. Sch\"{a}fer, H. C. D\"{o}nges, and U. Mosel, \PLB{342}{1995}{13}.

\bibitem{Alv73} H. Alvensleben, U. Becker, P. J. Biggs {\it et al.}, \PRL{30}{1973}{328}.

\bibitem{Lvo96} A. I. L'vov, S. Scopetta, D. Drechsel, and S. Scherer,
                                   in Ref.~\cite{Procee} p.186;\,  \PRC{57}{1998}{312}.
\bibitem{Kor97} A. Yu. Korchin, O. Scholten, and F. de Jong, \PLB{402}{1997}{1}.

\bibitem{Die97} A. E. L. Dieperink and S. I. Nagorny, \PLB{397}{1997}{20}.

\bibitem{Bac99} J. C. S. Bacelar, private communication.

 \bibitem{Bjo58} J. D. Bjorken, S. D. Drell, and S. C. Frautschi, \PR{112}{1958}{1409}.

\bibitem{Don86} T. W. Donnely, in {\it New Vistas in Electro-Nuclear Physics}, ed. by E. L. Tomusiak, H. S.
 Caplan, and E. T. Dressler (Plenum Press, New York, 1986), p.151.

 \bibitem{Sch70} K. Schilling, R. Seyboth, and G. Wolf, \NP{B15}{1970}{397}.

 \bibitem{Sch96} O. Scholten, A. Yu. Korchin, V. Pascalutsa, and D. Van Neck, \PLB{384}{1996}{13}.

 \bibitem{Kor98}  A. Yu. Korchin, O. Scholten, and R. G. E. Timmermans, \PLB{438}{1998}{1}.

 \bibitem{Itz80} C. Itzykson and J. -B. Zuber, {\it Quantum Field Theory} (McGraw-Hill Book Company, 1980).

 \bibitem{Nec94} D. Van Neck, A. E. L. Dieperink, and O. Scholten, \NP{A574}{1994}{643}.

 \bibitem{Kor98a} A. Yu. Korchin, D. Van Neck, M. Waroquier, O. Scholten, and A. E. L. Dieperink,
                                      \PLB{441}{1998}{17}.

 \bibitem{Kor95} A. Yu. Korchin and O. Scholten, \NP{A581}{1995}{493}.

 \bibitem{Mes} J. G. Messchendorp, J. C. S. Bacelar, M. J. van Goethem {\it et al.}, \PRL{83}{1999}{2530};\,
  Phys. Rev. C, in press.

 \bibitem{Dab67} J. Daboul, \NP{B4}{1967}{180}.

 \bibitem{Klo98} W. M. Kloet, Wen-Tai Chiang, and Frank Tabakin, \PRC{58}{1998}{1086}.

 \bibitem{Klo99} W. M. Kloet and Frank Tabakin, \PRC{61}{2000}{015501}.

 \bibitem{Pilk79} H. M. Pilkuhn, {\it Relativistic Particle Physics} (Springer-Verlag,
 New-York - Heidelberg - Berlin, 1979).

 \bibitem{Robs74} B. A. Robson, {\it The Theory of Polarization Phenomena} (Clarendon Press, Oxford, 1974).

 \bibitem{Pfe74} W. Pfeil, H. Rollnik, and S. Stankowski, \NP{B73}{1974}{166}.

 \bibitem{Dre92} D. Drechsel and L. Tiator, J. Phys. G: Nucl. Phys. {\bf 18}, 449 (1992).

 \bibitem{Pec69} R. D. Peccei, \PR{181}{1969}{1902}.

 \bibitem{Ols75} M. G. Olsson and E. T. Osypowski, \NP{B87}{1975}{399}.

 \bibitem{Ben89} M. Benmerrouche, R. M. Davidson, and N. C. Mukhopadhyay, \PRC{39}{1989}{2339}.

 \bibitem{Pas95} V. Pascalutsa and O. Scholten, \NP{A591}{1995}{658}.

 \bibitem{VMD} G. E. Brown, Mannque Rho, and W. Weise, \NP{A454}{1986}{669};\,
                            M. Gari and W. Krumpelmann, \ZPA{322}{1985}{689};\,
                            R. Williams, S. Krewald, and K. Linen, \PRC{51}{1995}{566}.

\bibitem{Lvo97} A. I. L'vov, V. A. Petrun'kin, and M. Schumacher, \PRC{55}{1997}{359};\,
 A. I. L'vov and A. M. Nathan, \PRC{59}{1999}{1064}, and private communication.
 In the calculation shown in~\figref{figu:2} the solution SM99K of the VPI
 analysis~\cite{SAID} has been used.

 \bibitem{SAID}  Virginia Tech SAID Facility, see {\tt http://said.phys.vt.edu/analysis};\, R. A. Arndt,
 I. I. Strakovskii, and R. L. Workman, \PRC{53}{1996}{430}.

 \bibitem{Hun97} A. H\"unger, J. Peise, A. Robbiano {\it et al.}, \NP{A620}{1997}{385}.        

 \bibitem{Dre00} D. Drechsel, M. Gorchtein, B. Pasquini, and M. Vanderhaeghen, \PRC{61}{2000}{015204}.

 \bibitem{Feu99} T. Feuster and U. Mosel, \PRC{59}{1999}{460}.


 \bibitem{Dew61} J. W. DeWire, M. Feldman, V. L. Highland, and R. Littauer, \PR{124}{1961}{909}. 

 \bibitem{Sti63} R. F. Stiening, E. Loh, and M. Deutsch, \PRL{10}{1963}{536}.                   

 \bibitem{Gen76} H. Genzel, M. Jung, R. Wedemeyer, and H. J. Weyer, \ZPA{279}{1976}{399}.      

 \bibitem{Tos78} K. Toshioka, M. Chiba, S. Kato {\it et al.}, \NP{B141}{1978}{364}.              

 \bibitem{Ish80} T. Ishii, K. Egawa, S. Kato {\it et al.}, \NP{B165}{1980}{189}.           

 \bibitem{Bar66} P. S. Baranov, V. A. Kuznetsova, L. I. Slovokhotov, G. A. Sokol, and L. N. Shtarkov,
                                Sov. J. Nucl. Phys. {\bf 3}, 791 (1966).               

 \bibitem{Bla96} G. Blanpied, M. Blecher, A. Caracappa {\it et al.}, \PRL{76}{1996}{1023}.  

 \bibitem{Ada93} F. V. Adamian, A. Yu. Buniatian, G. S. Frangulian {\it et al.}, J. Phys. G: Nucl. Part.
 Phys. {\bf 19}, L139 (1993).                                                                  

 \bibitem{Wad84} Y. Wada, K. Egawa, A. Imanishi, T. Ishii, S. Kato, K. Ukai, F. Naito, H. Hara,
                                           T. Noguchi, and K. Takahashi, \NP{B247}{1984}{313}.

 \bibitem{Bur96} V. D. Burkert, Czech. J. Phys. {\bf 46}, 627 (1996).

 \bibitem{Eis70} J. M. Eisenberg and W. Greiner, {\it Excitation mechanisms of the nucleus}
  (North-Holland Publishing Company, Amsterdam-London, 1970).

 \bibitem{Bij96} R. Bijker, F. Iachello, and A. Leviatan, \PRC{54}{1996}{1935}.

 \bibitem{Jac59} M. Jacob and G. C. Wick, \AP{7}{1959}{404}.

\end{thebibliography}
\end{document}